\documentclass[aps,prc,twocolumn,groupedaddress]{revtex4-2}

\usepackage{amsmath,amssymb,multirow}
\usepackage[dvipdfmx]{graphicx}
\usepackage{epsfig,amsfonts,amsbsy,color}
\usepackage{hyperref}
\def\Vec#1{\boldsymbol{#1}}

\newcommand{\pderf}[3]{\left(\frac{\partial #1}{\partial  #2}\right)_{#3}}

\newcommand{\der}[2]{\frac{d #1}{d  #2}}

\newcommand{\var}[2]{\frac{\delta #1}{\delta  #2}}
\newcommand{\bv}[1]{{\boldsymbol #1}}

\newcommand{\sbkt}[1]{\langle#1\rangle}

\newcommand{\ep}{\varepsilon}

\newcommand{\mc}{\mathrm{mc}}
\newcommand{\mcn}{\mathrm{ss}}
\newcommand{\can}{\mathrm{can}}
\newcommand{\bra}{\left\langle}
\newcommand{\ket}{\right\rangle}
\newcommand{\Tc}{T_{\mathrm{c}}}

\newcommand{\intTotx}{\int_0^{L_x}}

\begin{document}


\title{
Control of Metastable States by Heat Flux in the Hamiltonian Potts Model 
}


\author{Michikazu Kobayashi}
\affiliation{
  School of Environmental Science and Engineering, Kochi University of
  Technology, Miyanoguchi 185, Tosayamada, Kami, Kochi 782-8502, Japan
}

\author{Naoko Nakagawa}
\affiliation {Department of Physics,  
              Ibaraki University, Mito 310-8512, Japan}

\author{Shin-ichi Sasa}
\affiliation{Department of Physics, Kyoto University, Kyoto, 606-8502 Japan
}


\date{\today}

\begin{abstract}
The local equilibrium thermodynamics is a basic assumption of macroscopic descriptions of the out of equilibrium dynamics for Hamiltonian systems.
We numerically analyze the Hamiltonian Potts model
in two dimensions to study the violation of the assumption for phase
coexistence in heat conduction.
We observe that the temperature
of the interface between ordered and disordered states deviates
from the equilibrium transition temperature, indicating that
metastable states at equilibrium are stabilized by the influence of a heat
flux. We also find that the deviation is described by the formula
proposed in an extended framework of the thermodynamics.
\end{abstract}


\maketitle


{\em Introduction.--} 
The macroscopic dynamics of a Hamiltonian system driven by nonequilibrium
boundary conditions is  expected to be described by hydrodynamic equations
with local equilibrium thermodynamics \cite{Landau, deGroot, Zubarev,
Mclennan, Sasa, Hongo}.
However, there are exceptional cases, such as the shear flow 
near a liquid-gas critical point \cite{Onuki-Kawasaki}, where
suppression of critical fluctuations by the shear flow modifies
the thermodynamic properties \cite{Onuki}, and as a result, 
the local equilibrium thermodynamics is violated.
A natural question is whether the violation of the local
equilibrium thermodynamics occurs except at critical points.
The aim of this Letter is to provide a definitive example
by studying nonequilibrium dynamics near the first-order transition point.


A first-order phase transition is a different type of singularity
than the critical point \cite{FOT,FIFO,nano-VO2,nano-LG}.
A characteristic feature
of a first-order phase transition is the existence of
hysteresis \cite{hysteresis}. 
For an order-disorder transition \cite{ferroelectric,NLC},
the observed transition temperature when a material is cooled
from the disordered state is lower than the transition temperature
when the same material is heated from the ordered state.
Such transition temperatures, generally both in cooling and heating,
deviate from the equilibrium transition temperature $\Tc$.
Thus, the supercooled disordered or the superheated ordered states
are often observed as a transient dynamical process. These observations
may suggest that the metastable states become steady states when the
system sets up at nonequilibrium conditions. 
To explore this possibility, we consider steady
states in heat conduction where two heat baths with
temperatures $T_1$ and $T_2$ are attached at the left and right
sides of the system.


We assume that $T_1$ and $T_2$  satisfy $T_1 \le \Tc \le T_2$ to 
observe phase coexistence, where ordered
and disordered states appear at the low and high temperature sides, respectively, with 
a unique interface separating the two phases.
Our main question in this Letter is whether
the temperature of the interface 
is equal to $\Tc$. If the local equilibrium thermodynamics
is assumed to hold at each point of the system, the interface temperature
should be equal to the equilibrium transition temperature.
However, the validity
of this assumption is not obvious because of the existence of metastable
states. To our best knowledge,
there have been no experimental studies on this question,
while a rich variety of nonequilibrium phase-coexistence phenomena have been studied
including flow boiling heat transfer and  pattern formation in crystal growth 
\cite{boiling, crystal, Cannell,Zhong,Ahlers,Urban}.


We study the interface temperature by numerical
simulations of a model that exhibits phase coexistence in steady heat
conduction. Since there is no standard model for order-parameter
dynamics with conducting energy, in this Letter, we propose the Hamiltonian
Potts model in two dimensions.  We first confirm the coexistence of
ordered and disordered states in an isolated system
by numerically solving the Hamiltonian equation.
Then, imposing a heat flux at the boundary with the
total energy fixed, we produce the phase coexistence in steady
heat conduction. A remarkable property in this system is that
the interface temperature deviates from the equilibrium transition
temperature.
This indicates that metastable states are stabilized.
That is, metastability is controlled by the heat flux. 
Furthermore, we find that the deviation is well fitted
by the formula proposed in an extended framework of thermodynamics,
which we call {\it global thermodynamics}
\cite{Global-PRL, Global-JSP, Global-PRR}.
It provides a quantitative prediction
of the phase coexistence in the steady heat conduction,
in contrast to other extended frameworks of thermodynamics
\cite{Keizer,Eu,Jou,Oono-paniconi, Sasa-Tasaki,Bertin,Seifert-contact,Dickman}.

\begin{figure}[bt]
\begin{center}
\includegraphics[width=0.95\linewidth]{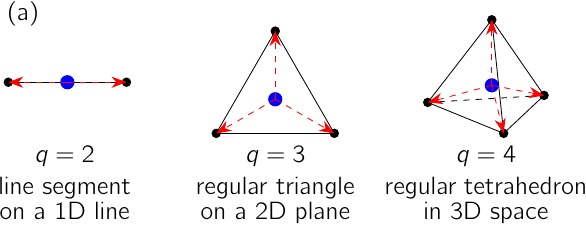} \\[10pt]
\includegraphics[width=0.9\linewidth]{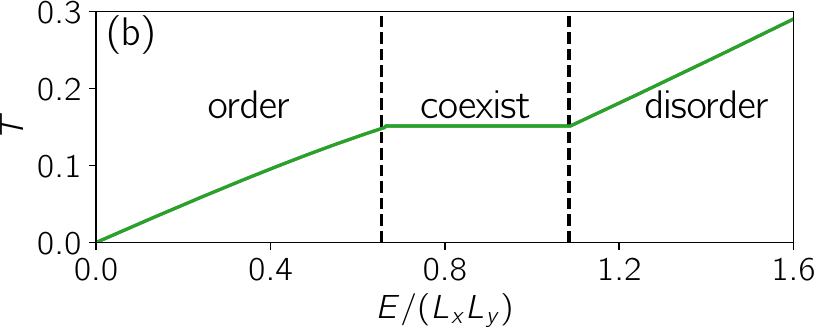}
\caption{
\label{fig:simplex}
(a) Examples of $(q-1)$ simplexes with $q$ vertices in $(q-1)$-dimensional
space: a line segment with $\mu_1$ and $\mu_2$ for $q=2$ (left), a regular
triangle with $\mu_1$, $\mu_2$ and $\mu_3$ for $q=3$ (middle),
and a regular tetrahedron with $\mu_1$, $\mu_2$, $\mu_3$, and $\mu_4$
for $q=4$ (right).
(b) Phase diagram for the $q=11$ Hamiltonian Potts model \eqref{Hamiltonian}.
The system shows the first-order phase transition at $T=\Tc$
{when $T$ is changed, while} the ordered and disordered states coexist 
in the range $E_1< E<E_2$. {$\Tc$, $E_1$, and $E_2$ are numerically
estimated} to be
$\Tc\simeq 0.15$, $E_1/(L_xL_y)\simeq 0.66$, and $E_2/(L_xL_y)\simeq 1.09$
\cite{SM}.}
\end{center}
\end{figure}

{\em Model and observables.--}
We study the Hamiltonian Potts model with $q$-fold symmetry in two dimensions.
Let $\phi^a(\bv{r})$, $a=1,2,\cdots q-1$, be a $q-1$ components field
defined in a rectangular region $D\equiv [0,L_x]\times[0,L_y]$
with $L_x>L_y$.
We express $(\phi^1, \cdots, \phi^{q-1})$ as $\phi$.  
The conjugate momentum field of $\phi^a(\bv{r})$ is expressed by $\pi^a(\bv{r})$,
where $\pi$ represents $(\pi^1, \cdots, \pi^{q-1})$. 
We assume the total Hamiltonian ${\cal H}$ as
\begin{equation}
{\cal H}(\phi,\pi) \!=\! \int_D d^2\bv{r}
\left\{ \frac{1}{2} \sum_{a=1}^{q-1}
\left[(\pi^a)^2 \!+\! \left|\bv{\nabla}\phi^a\right|^2\right]
\!+\! V(\phi) \right\},
\label{Hamiltonian}
\end{equation}
where the potential $V(\phi)$ possesses $q$ symmetric minima in
$\mathbb{R}^{q-1}$.
Let $\mu_k$, $1 \le k \le q$, be coordinates of the $q$ vertices for the regular ($q-1$) simplex in $\mathbb{R}^{q-1}$ as illustrated in
Fig. \ref{fig:simplex} (a) for the cases $q=2$, $3$, and $4$,
{where the centroid of the simplex is located  at the origin.}
The explicit form of $\mu_k$ is given in
Supplemental Material \cite{SM}.
As a potential with minima at $\mu_k$, we set  
\begin{equation}
V(\phi)=
\frac{1}{2}\prod_{k=1}^{q} \sum_{a=1}^{q-1}(\phi^a-\mu_k^a)^2.
\label{potential}
\end{equation}
This potential is regarded as a continuous extension of
the standard $q$-state Potts model.
We thus expect that the equilibrium statistical mechanics
for ${\cal H}$ describes the same phase transitions as those
observed in the standard Potts model. As for the standard
  $q$-state Potts model, the model exhibits a first-order
  transition at a temperature $\Tc$ for $q \ge 5$ in equilibrium
  \cite{Potts-review}.
In this Letter, we numerically study the case $q=11$ for $L_x=384$ and $L_y=64$.
The system is spatially discretized with a grid spacing
$\Delta x=1/8$ \cite{SM}.
The Boltzmann constant is set to unity.

We define the local kinetic energy density per unit degree of freedom 
and the local order parameter as
\begin{align}
\hat T(\bv{r}) =  \frac{\sum_{a=1}^{q-1}[\pi^a(\bv{r})]^2 }{q-1},\quad
\hat{m}(\Vec{r})\equiv\sum_{a=1}^{q-1}\phi^a(\bv{r})\mu_1^a,
\label{e:order-parameter}
\end{align}
{where the direction of symmetry breaking is fixed to the 
$\mu_1$ direction in our numerical simulations by choosing specific
initial conditions \cite{SM}.}
Their average over the $y$ direction with $x$ fixed {is expressed} as
\begin{align}
    [ \hat A ]_x=\frac{1}{L_y} \int_0^{L_y} dy\: \hat A(\bv{r}),
    \label{Ax}
\end{align}
where $\hat A$ is $\hat T$ or $\hat m$.

{\em Equilibrium phase coexistence.--} 
We first examine the equilibrium phase diagram by numerically investigating
the isothermal dynamics \cite{SM}.
The first-order transition is observed at $T=\Tc\simeq 0.15$, where
the energy density changes discontinuously at $T=\Tc$.
The system is occupied by the ordered and disordered states for $T<\Tc$ and $T>\Tc$ , respectively.
The phase diagram is shown in Fig. \ref{fig:simplex} (b).

With the knowledge of the phase diagram, we concentrate on an
isolated system with the total energy $E$ fixed.
The time evolution is given by  the Hamiltonian equation
\begin{align}
 \partial_t \phi^a =\var{{\cal H}}{\pi^a}, \quad
 \partial_t \pi^a = -\var{{\cal H}}{\phi^a}.
\label{Hamil} 
\end{align}
Note that 
\begin{equation}
  \der{{\cal H}}{t} = \int_D d^2\bv{r}
  ~\sum_{a=1}^{q-1} \bv{\nabla}\left( \pi^a \bv{\nabla} \phi^a \right).
\label{ene-evol}
\end{equation}
For conserving energy, we assume the Neumann boundary condition in the $x$ direction: 
\begin{equation}
 \left. \partial_x \phi \right|_{x=0,L_x} =0
\label{b-con}
\end{equation}
for any $y$, and periodic boundary conditions in the $y$ direction.

\begin{figure}[bt]
\centering
\includegraphics[width=0.95\linewidth]{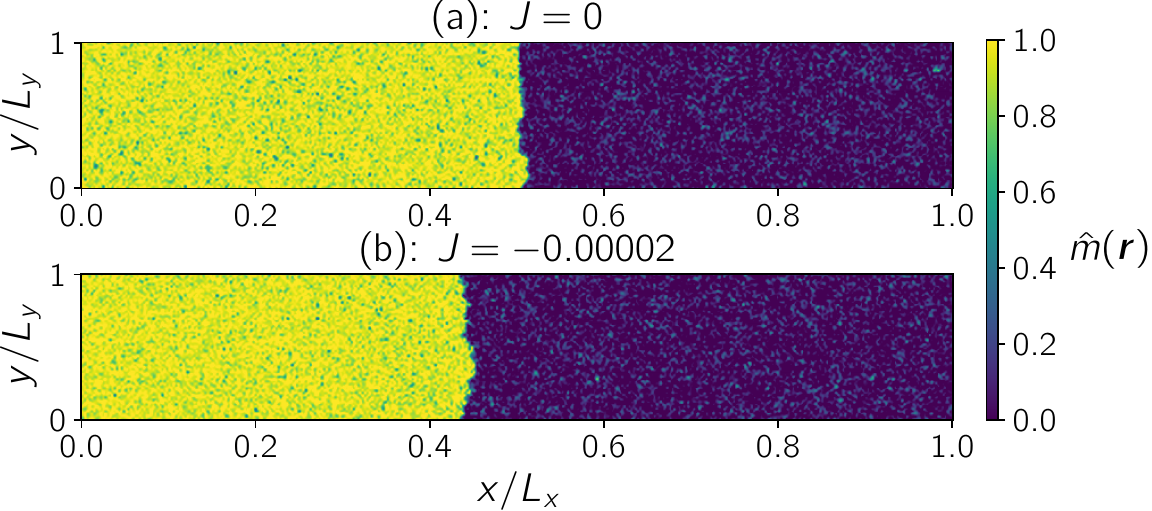}\\[5pt]
\includegraphics[width=0.95\linewidth]{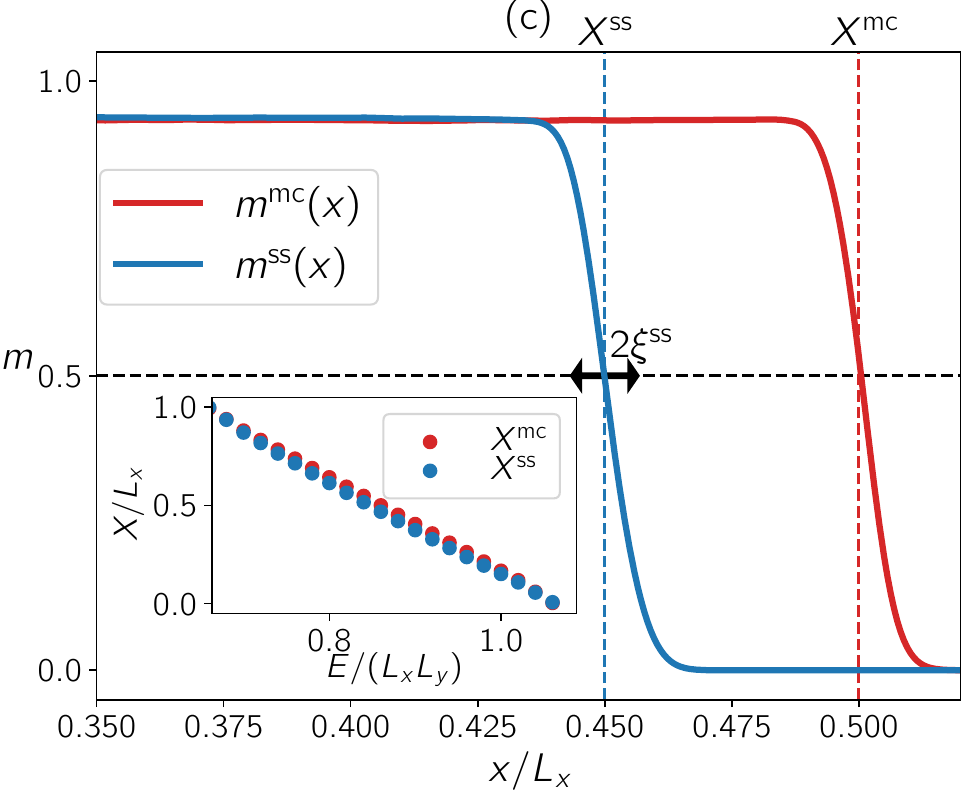}
\caption{
\label{fig:order-snapshot}
(a), (b) Snapshot of the order parameter density $\hat{m}(\Vec{r})$ in equilibrium with $J=0$ [panel (a)] and the steady state with $J=-0.00002$ [panel (b)]
for $E/(L_xL_y)=0.88$. 
(c) One-dimensional order parameter profiles $m^\mc(x)$ and $m^\mcn(x)$. 
The interface positions $X^\mc$ and $X^\mcn$ are obtained as $m^\mc(X^\mc)=0.5$ and $m^\mcn(X^\mcn)=0.5$, respectively.
The interface thickness $\xi^\mcn$ is estimated by using $m^{\rm ss}(x)
= a - b \tanh[(x-x_0)/\xi^\mcn]$.
Inset: Dependence of $X^{\mc}$ and $X^{\mcn}$ on the energy density $E/(L_xL_y)$.
}
\end{figure}

We start from initial conditions with a single interface {and solve the
Hamiltonian equation} \eqref{Hamil} 
until {an equilibrium state is realized} \cite{SM}.
Figure \ref{fig:order-snapshot}(a) shows a snapshot of
the local order parameter $\hat{m}(\Vec{r})$ in equilibrium
with an interface parallel to the $y$ axis minimizing the interfacial energy.
We thus discuss
the one-dimensional profile by using the average over $y$ defined by \eqref{Ax}.

Let $ \langle \hat A \rangle_E$ be the long-term
average of $\hat A(\bv{r})$ after equilibration, which corresponds to
the expected value with respect to the microcanonical distribution
with $E$.
We show the order parameter profile
$m^\mc(x)=\langle [\hat{m}]_x \rangle_E$ in Fig. \ref{fig:order-snapshot}(c), which exhibits a typical interface structure.
Here, the superscript ``mc" indicates ``microcanonical".
The interface position $X^\mc$
defined as $m^\mc(X^\mc)=0.5$ decreases approximately 
linearly with $E$  such that $X^\mc=L_x$ for $E=E_1$ and $X^\mc=0$ for $E=E_2$.
There is no interface for $E < E_1$ and $E > E_2$.

\begin{figure}[bt]
\begin{center}
\includegraphics[width=0.95\linewidth]{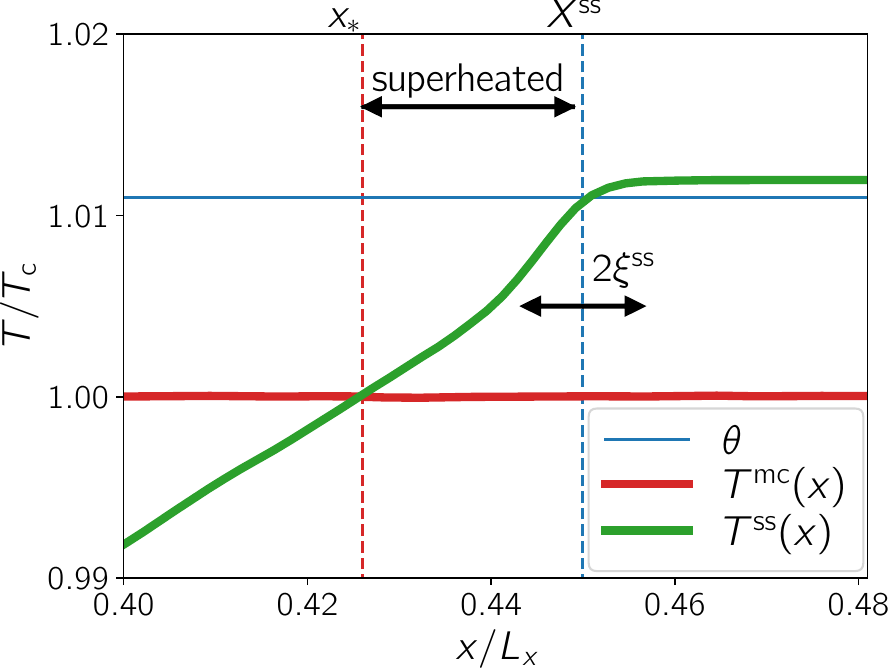}
\caption{
\label{fig:current}
One-dimensional temperature profile $T^\mcn(x)$ for $J=-0.00002$
and $T^\mc(x)$ for $J=0$, where  $E/(L_xL_y)=0.88$.
The blue dashed line represents the interface position $X^\mcn$.
The red dashed line represents the position $x_\ast$ for $T^\mcn(x_\ast)$ being the equilibrium transition temperature $\Tc$.
We note that the position $x_\ast$ is not at the interface
{region}, i.e., $x_\ast < X^\mcn - \xi^\mcn \simeq 0.444L$.
}
\end{center}
\end{figure}

The one-dimensional temperature profile
$T^\mc(x)=\langle[\hat T]_x\rangle_E$ is homogeneous in $x$ even for
the phase coexistence observed in $E_1<E<E_2$. 
This homogeneous temperature is equal to $\Tc$ obtained
in isothermal systems as shown in Fig. \ref{fig:current} \cite{SM}.
All these results show that the phase coexistence observed as an
equilibrium state for the case $E \in [E_1,E_2]$ is  an important
feature of the isolated system and that the behavior is equivalent to
the discontinuous change observed in the isothermal system, as displayed
in the phase diagram in Fig. \ref{fig:simplex} (b).

{\em Nonequilibrium phase coexistence.--}
We now consider phase coexistence in heat conduction.
Recall that the position of the interface is uniquely determined in
isolated systems for given $E\in [E_1,E_2]$,
whereas it is neutral and not under control in isothermal systems
at $T=\Tc$.
Thus, energy-conserving heat conduction could be preferable to standard heat
conduction for a detailed study of thermodynamic properties. 
From the thermodynamic equivalence for heat
conducting states as well as equilibrium states \cite{Global-PRR,SM}, 
we expect that the obtained results for the energy-conserving heat-conduction systems
will explain the phase coexistence observed in standard
heat-conducting systems.

We construct a heat-conducting system where the energy flows in at $x=L_x$ and flows out at $x=0$, while keeping the energy ${\cal H}$ constant.
That is, we impose a constant heat flux $(JL_y,0)$ at $x=0$ and $(JL_y,0)$ at $x=L_x$.
Recalling (\ref{ene-evol}), we set
\begin{align}
\left. \partial _x \phi^a \right|_{x=0,L_x} = -\frac{ \left. J L_y \pi^a \right|_{x=0,L_x}}{\int_0^{L_y} dy \left. \sum_{b=1}^{q-1} {(\pi^b)^2} \right|_{x=0,L_x}}, 
\label{b-con-noneq}
\end{align}
which is a nonequilibrium generalization of (\ref{b-con}).

We study the case $J=-0.00002$ \cite{fn:LR}. 
Figure \ref{fig:order-snapshot}(b) is a snapshot of the order-parameter
density field $\hat{m}(\Vec{r})$ {in the steady state, which} can be
compared with the equilibrium case shown in Fig. \ref{fig:order-snapshot}(a).
The interface in the steady state is shifted slightly to the left of
the equilibrium interface. 
In Fig. \ref{fig:order-snapshot}(c), we display the order-parameter profile
$m^\mcn(x) = \langle [\hat{m}]_x \rangle_{E,J}$,
where $\langle\cdot\rangle_{E,J}$ represents the long-term average
in the steady state with $E$ and $J$.
For the interface position $X^\mcn$ defined by $m^\mcn(X^\mcn)=0.5$,
we find that $|X^\mcn- X^\mc|/L \simeq 0.05$. The inset plots $X^\mcn$
for an energy density $E/(L_xL_y)$. The relative difference
$|X^\mcn- X^\mc|/L$ becomes smaller as $E$ approaches $E_1$ or $E_2$.

In Fig. \ref{fig:current}, we plot the temperature profile
$T^\mcn(x)=\langle [\hat{T}]_x \rangle_{E,J}$.
An important feature is that the position $x_\ast$ satisfying $T^\mcn(x_\ast)=\Tc$
is inside the ordered regions, i.e., $x_\ast < X^\mcn-\xi^\mcn$,
where $\xi^\mcn$ is the interface thickness as indicated in
Fig. \ref{fig:order-snapshot}(c).
That is,  superheated ordered states  appear
in a region where $T(x) > \Tc$ and $m(x) > 0.5$.
This means that the local equilibrium thermodynamics is
violated due to the heat flux. In other words, the temperature at the interface
$\theta\equiv T^\mcn(X^\mcn)$ deviates from $\Tc$.

\begin{figure}[bt]
\begin{center}
\includegraphics[width=0.95\linewidth]{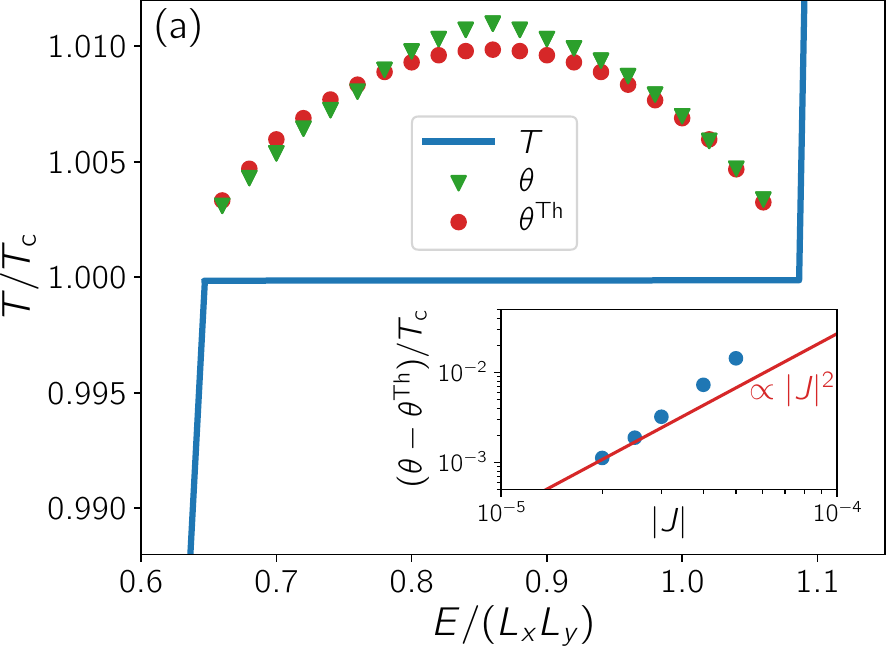} \\[10pt]
\includegraphics[width=0.95\linewidth]{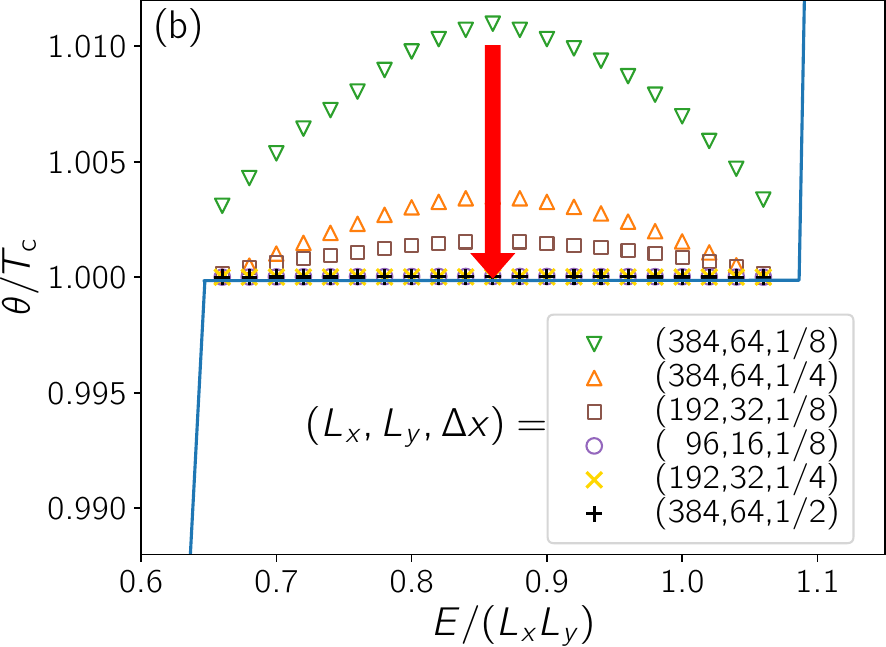}
\caption{
\label{fig:interface}
(a) 
Interface temperature as a function of the energy density $E/(L_xL_y)$ for $J=-0.00002$.
The green points are numerical values $\theta\equiv T^\mcn(X^\mcn)$ and the red points are the theoretical predictions $\theta^{\rm Th}$ in \eqref{e:theta-J}, substituting the numerically determined $X^\mc$ for $X$.
The blue line corresponds to the green line
  in Fig. \ref{fig:simplex} (b).
Inset: $\theta-\theta^{\rm Th}$ are plotted for $|J|$ at $E/(L_xL_y)=0.86$ that gives the peak value. $\theta-\theta^{\rm Th} $ approaches the red line representing a relationship proportional to $|J|^2$.
(b) Finite size effects of the numerically determined $\theta$ for $(L_x,L_y,\Delta x)$ depicted in the box with $J=-0.00002$. Several symbols are
overlapped \cite{SM}. }
\end{center}
\end{figure}

To study the violation more quantitatively, in Fig. \ref{fig:interface}(a),  
we plot $\theta$ as a function of $E$ with $J=-0.00002$ fixed. 
It is notable that the interface temperature $\theta$ deviates
from the equilibrium transition temperature $\Tc$ 
for $E \in [E_1,E_2]$.
{The deviation}
becomes a maximum around the midpoint of $[E_1,E_2]$, where the
interface position is around $L_x/2$.
The superheated regions
disappear for $E\rightarrow E_1$ or $E_2$, where the system becomes
occupied by an ordered or disordered state. 
To examine the {parameter} dependence,
we compare {the numerical results} for the violation with the formula
\begin{align}
\begin{aligned}
\theta^{\rm Th}=T_{\rm c}+{|J|}\left(\frac{1}{\kappa^{\rm o}}-\frac{1}{\kappa^{\rm d}}\right)\frac{X(L_x-X)}{2 L_x}, 
\end{aligned}
\label{e:theta-J}
\end{align}
which was proposed in 
the global thermodynamics framework \cite{Global-JSP},
where $\kappa^{\rm o}$ and $\kappa^{\rm d}$
are the heat conductivities of ordered and disordered states, respectively,
and $X$ is the position of the interface.
The formula \eqref{e:theta-J} can be derived for the order-disorder
transition \cite{SM}.
In Fig. \ref{fig:interface}(a), we simultaneously plot $\theta$ and $\theta^{\rm Th}$, where the latter is obtained by substituting
  $X^\mc$ into $X$ in \eqref{e:theta-J}.
We find that $\theta$ directly measured
in numerical simulations is in qualitative agreement with $\theta^{\rm Th}$
determined by the formula \eqref{e:theta-J}.
The discrepancy is due to a nonlinear effect of $|J|$
as shown in the inset of Fig. \ref{fig:interface} (a).

Finally, we investigate the finite size effects.
We find that the violation of local equilibrium becomes weaker
for smaller $L_x$ or larger $\Delta x$. Instead of
the original system consisting of $3073 \times 512$ grid points
with $\Delta x=1/8$, we study smaller systems consisting of $1537 \times 256$
grid points and $768 \times 128$ grid points, keeping the value of
$\Delta x$ fixed. The results are shown in Fig. \ref{fig:interface} (b).
The interface temperature deviates from the theoretical
curve and eventually reaches the equilibrium
transition temperature.
This indicates that long-wavelength fluctuations play an
important role in the violation of local equilibrium properties.  
A similar trend is observed for rough systems with larger values of $\Delta x$ for the same system size. The interface temperature becomes closer to $\Tc$
{as} $\Delta x$ increases from $1/8$ to $1/2$.
This implies that regularity of the short wavelength fluctuations
of the continuum fields is also necessary for the violation, which is
in stark contrast to the properties in equilibrium.
We do not observe the violation of local equilibrium with smaller
or rougher systems.
These observations
suggest that hydrodynamic fluctuations on some scales play an important role
for the violation of the local equilibrium at the interface.

{\em Concluding remarks.--}
In this Letter, we studied nonequilibrium phase
coexistence under a heat flux using the two-dimensional
Hamiltonian Potts model with $q=11$.
We found that a superheated ordered region stably appears in
the heat-conducting state. 
This indicates that the metastable states are controlled by the heat flux.
Our numerical results also quantitatively support the validity of
global thermodynamics. 
We hope this work triggers further experimental and theoretical
research, such as engineering  thermodynamic metastable states.
In conclusion, we present three remarks.

Although we expect that the violation of local equilibrium
thermodynamics is generically 
observed in other models and experiments, we need to
carefully choose the system conditions. For example,
liquid-gas coexistence under heat conduction was studied using
molecular dynamics simulations, and no violation was observed
\cite{Bedeaux,Yukawa}. We conjecture that the system sizes were too small
to allow long-wavelength fluctuations that would lead to the violation of local
equilibrium thermodynamics.
So far, we do not estimate a crossover system size beyond which
the violation is observed.
As a reference, we remark that a long-range correlation of hydrodynamic fluctuations in a sheared system is observed only when the particle number exceeds a crossover value of $10^7$ \cite{Nakano}.
We thus expect that the same order of particles are necessary for the
violation.  To confirm this conjecture is left for future study.

From a theoretical viewpoint, an important first study is to derive the stationary
distribution of the heat conduction system. In the linear response
regime, the Zubarev-McLennan representation is a generalization
of the microcanonical distribution, where the correction term is expressed
in terms of entropy production \cite{Zubarev,Mclennan}, as discussed
for phase coexistence in heat conduction \cite{Global-PRE}.
Developing a method for estimating this
correction term explicitly for microscopic or mesoscopic models is an important goal.

The most critical challenge is the observation of the violation
of local equilibrium thermodynamics in laboratory experiments.
As an example, for the liquid-gas coexistence for water at 1 atm
pressure, the interface temperature was calculated to be 96 $^\circ$C
when the temperatures of the heat baths attached to the left and
right sides were 95 and 105 $^\circ$C \cite{Global-JSP}.
This means that super-cooled gas stably appears near the interface
due to the heat flux. We believe that observation of this phenomenon
is of fundamental importance.


The authors thank F. Kagawa, K. Saito, and Y. Yamamura for useful discussions
on experiments of nonequilibrium phase coexistence; S. Yukawa and A. Yoshida for
discussions on molecular dynamics simulations of nonequilibrium phase
coexistence; and M. Itami and Y. Nakayama for discussions on fluctuating
hydrodynamics.
The work of M.K. was supported in part by the Osaka City University Advanced Mathematical Institute (MEXT Joint Usage/Research Center on Mathematics and Theoretical Physics, Grant No. JPMXP0619217849).
The present study was supported by JSPS KAKENHI Grants No.  JP19KK0066, No. JP20K03765, No. JP19K03647, No. JP19H05795, No. JP20K20425, and No. JP22H01144. 



\clearpage

\onecolumngrid

\setcounter{figure}{0}
\def\thefigure{S.\arabic{figure}}
\setcounter{equation}{0}
\renewcommand{\theequation}{S.\arabic{equation}}

\begin{center}
{\large \bf Supplemental Material for  \protect \\ 
  ``Control of metastable states by heat flux in the Hamiltonian Potts model'' }\\
\vspace*{0.3cm}
Michikazu Kobayashi$^{1}$, Naoko Nakagawa$^{2}$, and Shin-ichi Sasa$^{3}$
\\
\vspace*{0.1cm}
$^{1}${\small \it
School of Environmental Science and Engineering, Kochi University of Technology, Miyanoguchi 185, Tosayamada, Kami, Kochi 782-8502, Japan} \\
$^{2}${\small \it Department of Physics,
Ibaraki University, Mito 310-8512, Japan} \\
$^{3}${\small \it Department of Physics, Kyoto University,
Kyoto, 606-8502 Japan} 
\end{center}

This Supplemental Material consists of three sections. 
In Sec. \ref{sec:global}, we review global thermodynamics for the
system we study. The goal of this section is to derive the
formula \eqref{e:theta-J} in the main text. In Sec. \ref{sec:details},
we present details of numerical simulations of the Hamiltonian equation
\eqref{Hamil} to  support the results of the main text. In Sec. \ref{sec:isothermal},
we explain the isothermal dynamics by which the equilibrium phase diagram
for the temperature is determined.

\section{Global thermodynamics}
\label{sec:global}

First of all, we introduce an external field $H\ge 0$ for formulating thermodynamics of ordered and disordered states. While
  the numerical computation in the main text was performed without imposing any external field, we study a spontaneous symmetry-breaking state
  by considering  the limit $H\rightarrow 0^+$. 

To set up the system, we notice the order parameter field $\hat m(\bv{r})$ defined as \eqref{e:order-parameter}
and introduce a uniform field $H$ acting on $\hat m(\bv{r})$ by
assuming the Hamiltonian of the system as
\begin{equation}
{\cal H}(\phi,\pi)- H {\hat M}(\phi).
\label{e:H_H}
\end{equation}
The Hamiltonian $\cal H$ without the external field 
is defined in \eqref{Hamiltonian}
and the total order parameter $\hat M$ is defined as the spatial integration
\begin{equation}
\hat M = \int d^2 \bv{r}~ \hat m(\bv{r})  .
\end{equation}
We assume that thermodynamic properties are not qualitatively changed by applying the external field for sufficiently small $H$. For instance, the equilibrium transition temperature may continuously change with $H$, which we write $\Tc(H)$. 
Below $\Tc(H)$ the ordered phase appears as studied in the main text. From the symmetry, $\Tc(H)=\Tc(0)+O(H^2)$.

In Sec. \ref{sec:eq-th}, we briefly review equilibrium thermodynamics.
In Sec. \ref{sec:bulk-th}, we formulate global thermodynamics for the bulk of ordered and disordered states in steady heat conduction.
In Sec. \ref{sec:coex-th}, we present the variational principle for determining the steady phase coexistence state in heat conduction.
Section \ref{sec:deriv-theta-J} is devoted to the derivation of \eqref{e:theta-J} used in the main text.
In Sec. \ref{sec:Legendre}, we explain the thermodynamic relations in heat conduction and the equivalence of steady states of the energy-conserved system with the system conducted by two heat baths.

\subsection{Equilibrium thermodynamics}
\label{sec:eq-th}

We review equilibrium thermodynamics for the model we study in
the main text. 

In the thermodynamic limit, $\cal H$ and $\hat M$ form the internal energy $U$ and macroscopic order parameter $M$.
Letting $E$ be the total energy of the system, \eqref{e:H_H} leads to $E=U-HM$.
Changing the value of $H$ to $H+dH$ requires work $d'W=-MdH$. 
Since $d'Q=TdS$, the energy conservation forms the first law of thermodynamics as
\begin{align}
dE=TdS-MdH.
\end{align}
Thus, the entropy $S$ is given as a function of $(E,H)$ for
a fixed volume of the system $V=L_xL_y$ with the relation
\begin{align}
dS=\frac{dE}{T}+\frac{M}{T}dH.
\end{align}
Since $S(E,H,V)$ satisfies
the extensivity 
\begin{align}
S(\lambda E, H, \lambda V)=\lambda S(E, H, V) 
\end{align}
for any $\lambda >0$, we can derive 
\begin{align}
dS=\frac{dE}{T}+\frac{M}{T}dH-\frac{f}{T}dV
\end{align}
with $f=(E-TS)/V$ which corresponds to the free-energy density.
Since we study the case $H=0$ in the main text, the internal energy $U$ is equal to the total energy $E$.

\subsection{Heat conduction without phase coexistence}
\label{sec:bulk-th}

Next, we study the linear response regime in heat conduction
systems without phase coexistence. Specifically, we consider
the case that the heat flux is in parallel to the $x$-axis
for simplicity. We then assume the existence of local thermodynamic quantities $a(\bv{r})$, which are independent of $y$.
We express them as $a(x)$ simply. 
The arguments below follow Sec. 3 of ref. \cite{Global-JSP-sm}.

Let $s(x)$, $T(x)$, $e(x)$, and $m(x)$ be local entropy density,
local temperature, local energy density, and local order parameter density. 
These local quantities satisfy the fundamental relation,
\begin{align}
    ds(x)=\frac{de(x)}{T(x)}+\frac{{m}(x)}{T(x)} d{H},
    \label{e:ds}
\end{align}
for each $x$. The total amounts of extensive quantities in heat conduction
are obtained by the integration of corresponding local quantities such that
\begin{align}
  S=L_y\intTotx dx ~s(x), \quad E=L_y\intTotx dx~e(x),
  \quad M=L_y\intTotx dx~m(x).
\label{e:globalS}
\end{align}
Hereafter, we call these ``global thermodynamic quantities."
We then define a global temperature and a global free energy density to characterize the heat conduction states as the mean value of the respective local values as
\begin{align}
    \tilde T=\frac{1}{L_x}\intTotx dx~ T(x),\label{e:globalT}\\
    \tilde f=\frac{1}{L_x}\intTotx dx~ f(x).\label{e:globalf}
\end{align}
In the linear response regime, the quantities defined by the spatial
integration such as \eqref{e:globalS}, \eqref{e:globalT} and \eqref{e:globalf} are estimated by the trapezoidal
rule. Thus the global quantities are represented by
corresponding local thermodynamic quantities at the midpoint $x=L_x/2$.
This indicates that the nonequilibrium entropy
in \eqref{e:globalS} is approximated by the equilibrium entropy function; i.e.,
\begin{align}
    S=S(E,H,V)+O(\ep^2),
\end{align}
where $\ep$ is a dimensionless small parameter proportional to $J$.
Straightforwardly, the global thermodynamic quantities in heat conduction
satisfy the fundamental relation of thermodynamics
\begin{align}
  dS=\frac{dE}{\tilde T}+\frac{M}{\tilde T} dH
   -\frac{\tilde f}{\tilde T}dV,
    \label{e:dS-neq}
\end{align}
with an error of $O(\ep^2)$.
While global quantities $E$, $M$, and $V$ are naturally defined
for the heat conduction states, $\tilde T$ and $\tilde f$ are given
as nonequilibrium extensions of the temperature and the free energy density
through the thermodynamic relation \eqref{e:dS-neq}.

\subsection{Phase coexistence in heat conduction}
\label{sec:coex-th}


Let us concentrate on the energy-conserving heat-conduction system. 
Choosing the value of $E$ in a certain range, say $[E_1,E_2]$,
we observe the phase coexistence with a planner interface. 
In equilibrium, the temperature at the phase coexistence is kept at the transition temperature $\Tc(H)$.
Now, we impose the energy flux $J$ in such a way that
  the total energy $E$ is conserved, where $J<0$ is assumed without
loss of generality. We then observe the phase coexistence of
the ordered state in $0 \le x \le X$ and the disordered state in
$X \le x \le L_x$, where $X$ is to the interface position.
The temperature field $T(x)$ and the order parameter field $m(x)$
satisfy the local thermodynamic relation in each phase, while
there is no established law for determining the interface temperature
$\theta$. Here, let us recall that the equilibrium state in the
phase coexistence for given $(E, H, V)$ is determined by the
maximum entropy principle, where the equilibrium state maximizes
the total entropy. We extend this variational principle to that
for heat conduction states. 

We start with global thermodynamics in each region introduced in Sec. \ref{sec:bulk-th}. 
The global entropies are expressed as
\begin{align}
    S^{\rm o}=S(E^{\rm o}, H, L_yX),\quad S^{\rm d}=S(E^{\rm d}, H, L_y(L_x-X)),
\end{align}
where we put the superscript ``o" or ``d" to indicate the quantities
in the ordered or disordered phase.
From the fundamental relation
\eqref{e:dS-neq}, we  have
\begin{align}
&\pderf{S^{\rm o}}{E^{\rm o}}{H,X}=\frac{1}{\tilde T^{\rm o}},\quad
\pderf{S^{\rm d}}{E^{\rm d}}{H,L-X}=\frac{1}{\tilde T^{\rm d}},\\
&\frac{1}{L_y}\pderf{S^{\rm o}}{X}{E^{\rm o},H}
=-\frac{\tilde f^{\rm o}}{\tilde T^{\rm o}},\quad
\frac{1}{L_y}\pderf{S^{\rm d}}{(L_x-X)}{E^{\rm d},H}
=-\frac{\tilde f^{\rm d}}{\tilde T^{\rm d}}.
\end{align}
Letting $T_1=T(0)$, $T_2=T(L)$, and $\theta=T(X)$, the trapezoidal rule yields
\begin{align}
\tilde T^{\rm o}=\frac{T_1+\theta}{2}, \quad
\tilde T^{\rm d}=\frac{T_2+\theta}{2}.
\label{e:meanT}
\end{align}
Note that $\theta$ is the interface temperature. 
$T_1$ and $T_2$ are temperatures at the left and right
boundaries of the system, respectively. 

For determining steady states in phase coexistence, thermodynamics provides variational principles.
For the equilibrium case $J=0$,  the equilibrium state with $E$, $H$, and $V$ fixed maximizes the total entropy
$S=S^{\rm o}+S^{\rm d}$ with respect to $E^{\rm o}$ and $X$.
Here, $E$ is the total amount of
energy of the system and $E^{\rm d}=E-E^{\rm o}$.  The variational
equations lead to $\tilde T^{\rm o}=\tilde T^{\rm d}$ and
$\tilde f^{\rm o}=\tilde f^{\rm d}$. 
For heat conduction cases
$J <0$, as derived in \cite{Global-PRR-sm}, the entropy as the
variational function is concluded as
\begin{align}
    & S=S^{\rm o}+S^{\rm d}+\phi\Psi
    \label{e:globalsS-neq}
\end{align}
with
\begin{align}
   & \phi=\frac{T_2-T_1}{\Tc(H)},\label{e:phi}\\
   & \Psi=\frac{E^{\rm d}X-E^{\rm o}(L_x-X)}{2 \Tc(H)L_x}. \label{e:Psi}
\end{align}
The steady state with $E$, $H$, $L_x$, and $\phi$ fixed is determined as the state maximizing
\eqref{e:globalsS-neq} with respect to $E^{\rm o}$ and $X$.
We then obtain the condition for the steady state as
\begin{align}
    & \tilde T^{\rm d}-\tilde T^{\rm o}=\frac{\phi \Tc(H)}{2}, \label{e:var1}\\
    & \frac{\tilde f^{\rm o}}{\tilde T^{\rm o}}-\frac{\tilde f^{\rm d}}{\tilde T^{\rm d}}=\phi\frac{E}{2\Tc(H)L_xL_y}. \label{e:var2}
\end{align}
Note that these two relations are the extension of the equilibrium balance relations $\tilde T^{\rm o}=\tilde T^{\rm d}$ and $\tilde f^{\rm o}=\tilde f^{\rm d}$.
Due to the nonequilibrium contributions in the right-hand sides of \eqref{e:var1} and \eqref{e:var2}, we find that the steady states contain metastable states in a region adjacent to the interface.  See \eqref{e:theta-J} and its derivation in Sec. \ref{sec:deriv-theta-J}.

\subsection{Derivation of \eqref{e:theta-J}}
\label{sec:deriv-theta-J}

For simplicity of the derivation below, we introduce $\alpha\equiv -f/T$.
Since $\alpha=s-e/T$, the equilibrium fundamental relation \eqref{e:ds} yields
\begin{align}
d\alpha=\frac{e}{T^2}dT+\frac{m}{T}dH.
\label{e:d-alpha}
\end{align}
We also notice $E=E^{\rm o}+E^{\rm d}$,
$E^{\rm o}=e_{\rm c}^{\rm o}XL_y+O(\ep)$,
and $E^{\rm d}=e_{\rm c}^{\rm d}(L_x-X)L_y+O(\ep)$
with $e_{\rm c}^{\rm o/d}\equiv e^{\rm o/d}(T_{\rm c}(H), H)$. 
Using $\alpha$ and  these relations, \eqref{e:var2} is written as
\begin{align}
\tilde \alpha^{\rm d}-\tilde \alpha^{\rm o}=\frac{\phi}{2\Tc(H)}
\left(e_{\rm c}^{\rm o}\frac{X}{L_x}
+e_{\rm c}^{\rm d}\frac{L_x-X}{L_x}\right)+O(\ep^2).
\label{e:var2-2}
\end{align}
Since $\tilde \alpha^{\rm o/d}=\alpha^{\rm o/d}(\tilde T^{\rm o/d},H)$
with the equilibrium functions $\alpha^{\rm o/d}(T,H)$,
we have
\begin{align}
\tilde\alpha^{\rm d}-\tilde\alpha^{\rm o}
=\alpha^{\rm d}(\Tc(H),H)+\frac{e_{\rm c}^{\rm d}}{\Tc(H)^2}(\tilde T^{\rm d}-\Tc(H))
-\alpha^{\rm o}(\Tc(H),H)-\frac{e_{\rm c}^{\rm o}}{\Tc(H)^2}(\tilde T^{\rm o}-\Tc(H))+O(\ep^2).
\end{align}
Substituting this into \eqref{e:var2-2}
and using the equilibrium balance
$\alpha^{\rm o}(\Tc(H),H)=\alpha^{\rm d}(\Tc(H),H)$,  
we obtain
\begin{align}
  e_{\rm c}^{\rm o}\left(\tilde T^{\rm o}-\Tc(H)
  +\frac{\phi \Tc(H)}{2}\frac{X}{L_x}\right)
-
e_{\rm c}^{\rm d}\left(\tilde T^{\rm d}-\Tc(H)
-\frac{\phi \Tc(H)}{2}\frac{L_x-X}{L_x}\right)
=0.
\label{e:var2-3}
\end{align}

The global temperature $\tilde T$ for the phase coexistence,
which is defined by \eqref{e:globalT}, is expressed as
\begin{align}
\tilde T=\tilde T^{\rm o}\frac{X}{L_x}+\tilde T^{\rm d}\frac{L_x-X}{L_x}.
\end{align}
Combining this with \eqref{e:var1}, we have
\begin{align}
\tilde T^{\rm o}=\tilde T-\frac{\phi \Tc(H)}{2}\frac{L_x-X}{L_x},\quad
\tilde T^{\rm d}=\tilde T+\frac{\phi \Tc(H)}{2}\frac{X}{L_x},
\label{e:To-Td}
\end{align}
where we have used $\tilde T=\Tc(H)+O(\ep)$.
Substituting \eqref{e:To-Td} into \eqref{e:var2-3}, we obtain
\begin{align}
(e_{\rm c}^{\rm d}-e_{\rm c}^{\rm o})
\left(\tilde T-\Tc(H)-\frac{\phi \Tc(H)}{2}\frac{L_x-2X}{L_x}
\right)=0  .
\label{e:var2-4}
\end{align}
Note that $e_{\rm c}^{\rm d}-e_{\rm c}^{\rm o}\neq 0$ in the first-order
transition. 
The relation \eqref{e:var2-4} together with \eqref{e:To-Td} leads to
\begin{align}
\tilde T^{\rm o}=\Tc(H)-\frac{\phi \Tc(H)}{2}\frac{X}{L_x}, \quad
\tilde T^{\rm d}=\Tc(H)+\frac{\phi \Tc(H)}{2}\frac{L_x-X}{L_x}.
\label{e:To-Td-var}
\end{align}
Applying \eqref{e:meanT}, the first relation of \eqref{e:To-Td-var}
is rewritten as
\begin{align}
T_1=2\Tc(H)-\theta-\phi \Tc(H)\frac{X}{L_x}.
\label{e:T1-var}
\end{align}

Finally, let us recall that the heat flux is
uniform in $x$ from the conservation of energy, which is expressed as 
\begin{align}
J=-\kappa^{\rm o}\frac{\theta-T_1}{X}=-\kappa^{\rm d}\frac{T_2-\theta}{L_x-X},
\label{e:heat-current}
\end{align}
where $\kappa^{\rm o/d}$ is the heat conductivity at the transition temperature, $\kappa^{\rm o/d}=\kappa^{\rm o/d}(\Tc(H),H)$.
\eqref{e:heat-current} is transformed to
\begin{align}
&T_1=\theta+J\frac{X}{\kappa^{\rm o}},\\
  &\phi=-\frac{J}{T_{\rm c}}\left(\frac{X}{\kappa^{\rm o}}
  +\frac{L_x-X}{\kappa^{\rm d}}\right).
\end{align}
We now substitute these two relations into \eqref{e:T1-var}. We then obtain
\begin{align}
  \theta=\Tc(H)-J
  \left(\frac{1}{\kappa^{\rm o}}-\frac{1}{\kappa^{\rm d}}\right)
  \frac{X(L_x-X)}{2L_x}.
  \label{eq:theta-prediction}
\end{align}
To this point, we have assumed $J<0$. Repeating the same argument
for the case $J>0$, we find that $\theta$ is symmetric for 
  the transformation $J \to -J$.  We thus conclude \eqref{e:theta-J}.

\subsection{Thermodynamic relations in phase coexistence}
\label{sec:Legendre}

Thermodynamic entropy corresponds to the maximum value of the variational function \eqref{e:globalsS-neq}.
With $\phi$ and $\Psi$ given in \eqref{e:phi} and \eqref{e:Psi}, and applying 
the variational equations \eqref{e:var1} and \eqref{e:var2}, we have the fundamental relation of global thermodynamics as \cite{Global-PRR-sm}
\begin{align}
    dS=\frac{dE}{\tilde T}+\frac{M}{\tilde T} dH-\frac{\tilde f}{\tilde T}dV+\Psi d\phi.
    \label{e:dS-neq-coexist}
\end{align}
Thus, the global entropy is $S=S(E,H,V,\phi)$, i.e., a different function from the equilibrium entropy.

As a more familiar setup for heat conduction, one may consider a system between two heat baths of $T_1$ and $T_2$.
Since the total energy $E$ is not conserved, natural thermodynamic variables would be $(\tilde T,H,V,\phi)$ considering a straightforward extension from
that for  equilibrium systems with a heat bath.
Here the global temperature $\tilde T$ is defined by \eqref{e:globalT} and connected to the global temperatures in the ordered and disordered regions as
\begin{align}
\tilde T=\frac{X}{L_x}\tilde T^{\rm o}+\frac{L_x-X}{L_x}\tilde T^{\rm d}.
\label{e:globalT-ToTd}
\end{align}
We can show the equivalence of the steady state in heat conduction between such systems and the energy-conserved systems investigated in the main text. 
More precisely, \eqref{e:theta-J} results from the minimization of a variational function corresponding to the spatial integration of free energy $F=L_y\int_0^{L_x} f(x) dx$. It is written as $F=F^{\rm o}+F^{\rm d}$ with $F^{\rm o}=F(\tilde T^{\rm o},H,L_y X)$ and $F^{\rm d}=F(\tilde T^{\rm d},H,L_y (L_x-X))$, and
the minimum value of $F$ satisfies a thermodynamic relation
\begin{align}
dF=-Sd\tilde T-MdH+\tilde f dV-\tilde T\Psi d\phi,
\label{e:dF-neq-coexist}
\end{align}
which indicates the nonequilibrium free energy is written as $F=F(\tilde T,H,V,\phi)$.
The two fundamental relations \eqref{e:dS-neq-coexist} and \eqref{e:dF-neq-coexist} are connected by the Legendre transformation with
\begin{align}
F=E-\tilde T S+O(\ep^2),
\end{align}
which is led from the spatial integration of the local thermodynamic relation
\begin{align}
f(x)=e(x)-T(x)s(x)
\end{align}
using $\tilde T$ in \eqref{e:globalT-ToTd} and the entropy in \eqref{e:globalsS-neq}.


\section{Details of  numerical simulations}
\label{sec:details}

\subsection{Explicit form for $\mu_k$ of \eqref{potential}}

We explicitly derive coordinates
$\mu_k^a$ ($1 \leq k \leq q$, $1 \leq a \leq q-1$) of the $q$ vertex for the regular $(q-1)$ simplex
in $\mathbb{R}^{q-1}$. We first notice that
$\mu_k^a$ should satisfy the following three conditions: \\
(i) the centroid is located
at the origin:
\begin{align}
\sum_{k=1}^{q} \mu_k^a = 0,
\label{con1}
\end{align}
(ii) the length between the centroid and each vertex is one: 
\begin{align}
\sum_{a=1}^{q-1} (\mu_k^a)^2 = 1,
\label{con2}
\end{align}
(iii) the length between two different vertices is constant:
\begin{align}
  \sum_{a=1}^{q-1} (\mu_k^a - \mu_{k^\prime}^a)^2
  = K (1-\delta_{kk^\prime})
\label{con3}
\end{align}
for any $1 \leq k,k^\prime \leq q$, where $K$ is a constant.
We then find that \eqref{con1}, \eqref{con2}, and \eqref{con3} are
satisfied by
\begin{align}
    \mu_k^a = \begin{cases}
    \displaystyle
    \frac{\sqrt{q}}{\sqrt{q-1}} \delta_k^a - \frac{\sqrt{q}+1}{(q-1)^{3/2}}, & 1 \leq k \leq q-1, \\
    \displaystyle
    \frac{1}{\sqrt{q-1}}, & k=q,  \\
    \end{cases}
    \label{eq:simplex-coordinate}
\end{align}
with $K=2q/(q-1)$.

\subsection{Numerical simulations of the Hamiltonian equation \eqref{Hamil} with the boundary condition \eqref{b-con}}
\label{sec:mcan}

  We numerically solve the Hamiltonian equation \eqref{Hamil}.
  By imposing the boundary condition \eqref{b-con}, the system conserves
  the total amount of energy $E$.
  Hereafter, we consider only the case with odd $q$.

First, we discretize a box of $(L_x,L_y)=(384,64)$ with a lattice spacing length $\Delta x=1/8$ in both $x$ and $y$ directions. Then the system consists of $N_x\times N_y$ grid points, i.e., $(N_x,N_y)=(3073,512)$.
To calculate the spatial derivative $\nabla^2 \phi^a$ included in ${\delta \cal H}/\delta \phi^a$, we use the following spectral collocation method \cite{Fornberg-sm}.
We first construct $(q-1)/2$-complex fields
\begin{align}
\Phi^{a}=\phi^{2a-1}+i\phi^{2a}
\label{e:Phi-phi}
\end{align}
from $(q-1)$-real fields $\phi^a$, where $1 \le a \le (q-1)/2$.
We evaluate the spatial derivative
$\nabla^2 \Phi^a=\nabla^2\phi^{2a-1}+i\nabla^2 \phi^{2a}$
by the Fourier transformation
\begin{align}
\begin{aligned}
&
  \nabla^2\Phi^a(x,y,t)
  =-\frac{1}{(N_x-1)N_y}\sum_{n_x=0}^{N_x-1}{}^\prime
  \sum_{n_y=-N_y/2+1}^{N_y/2}(k_x^2 + k_y^2)
  \tilde{\Phi}^a(n_x,n_y,t)\cos(k_xx)e^{ik_yy}, \\
&
k_x = \frac{\pi n_x}{L_x}, \quad
k_y = \frac{2 \pi n_y}{L_y},
\end{aligned}
\label{eq:mcan-Laplacian}
\end{align}
where we define
\begin{align}
  \sum_{n=n_i}^{n_f}{}^\prime
  f(n)\equiv \frac{f(n_i)+f(n_f)}{2}+\sum_{n=n_i+1}^{n_f-1}f(n).
  \label{e:sum-prime}
\end{align}
The Fourier basis functions in \eqref{eq:mcan-Laplacian} are
cosine functions in $x$ direction to satisfy the boundary condition \eqref{b-con}.
The Fourier component $\tilde{\Phi}^a(n_x,n_y,t)$
in \eqref{eq:mcan-Laplacian} is obtained by the inverse transformation
\begin{align}
\begin{aligned}
  \tilde{\Phi}^a(n_x,n_y,t) =
  2 \sum_{I=0}^{N_x-1}{}^\prime
  \sum_{J=0}^{N_y-1} \Phi^a(x,y,t) \cos(k_xI\Delta x)e^{-ik_yJ\Delta x},
\end{aligned}
\end{align}
where $\Delta x$ is the lattice spacing length.
The variables $\Phi$ defined on  the $N_x\times N_y$ grid points
  evolve according to the Hamiltonian equation \eqref{Hamil}. 
To obtain the time evolution of them, we adopt
the 4th-order Strang splitting (symplectic integrator) method \cite{McLachlan-sm} with the temporal interval $\Delta t=1/4096$.
We confirmed that the energy is conserved as $|\bra\mathcal{H}\ket_t-\bra\mathcal{H}\ket_{t=0}|/\bra\mathcal{H}\ket_{t=0}<10^{-8}$.

We denote the solution of the Hamiltonian equation \eqref{Hamil}
with the boundary condition \eqref{b-con} as $(\phi,\pi)_E(\Vec{r},t)$,
where $E$ is the amount of the energy (the value of $\mathcal{H}$), which is determined by the initial condition and conserved in the time evolution. 
Aiming at observing the phase coexistence of the ordered and disordered states, we set the initial condition 
as a connected configuration of
two equilibrium configurations prepared at different temperatures:
\begin{align}
  (\phi,\pi)_E(x,y,0)=
  \begin{cases}
    (\phi_1,\pi_1)_{T_1}(x,y,t_{\rm f}) & 0\leq x<L_x/2, \\
    (\phi_1,\pi_1)_{T_2}(x-L_x/2,y,t_{\rm f}) & L_x/2\leq x<L_x, \\
    (\phi_1,\pi_1)_{T_2}(0,y,t_{\rm f}) & x=L_x,
  \end{cases}
    \label{eq:microcanonical-initial}
\end{align}
where $(\phi_1,\pi_1)_{T}$ is the solution of 
\eqref{eq:Ohzeki-Ichiki-under} with the isothermal condition at $T$, which will be explained in Sec. \ref{sec:isothermal}.
Thus, the energy $E$ is determined depending on the chosen initial
condition with $T_1$ and $T_2$.
We fix $T_1=0.63\Tc$, and change $T_2$ from $0.63 \Tc$ to $2.0 \Tc$.
Accordingly,  the energy $E$ changes from $E/(L_xL_y)=0.5$ to $E/(L_xL_y)=1.2$. 

\begin{figure}[bt]
\centering
\includegraphics[width=0.9\linewidth]{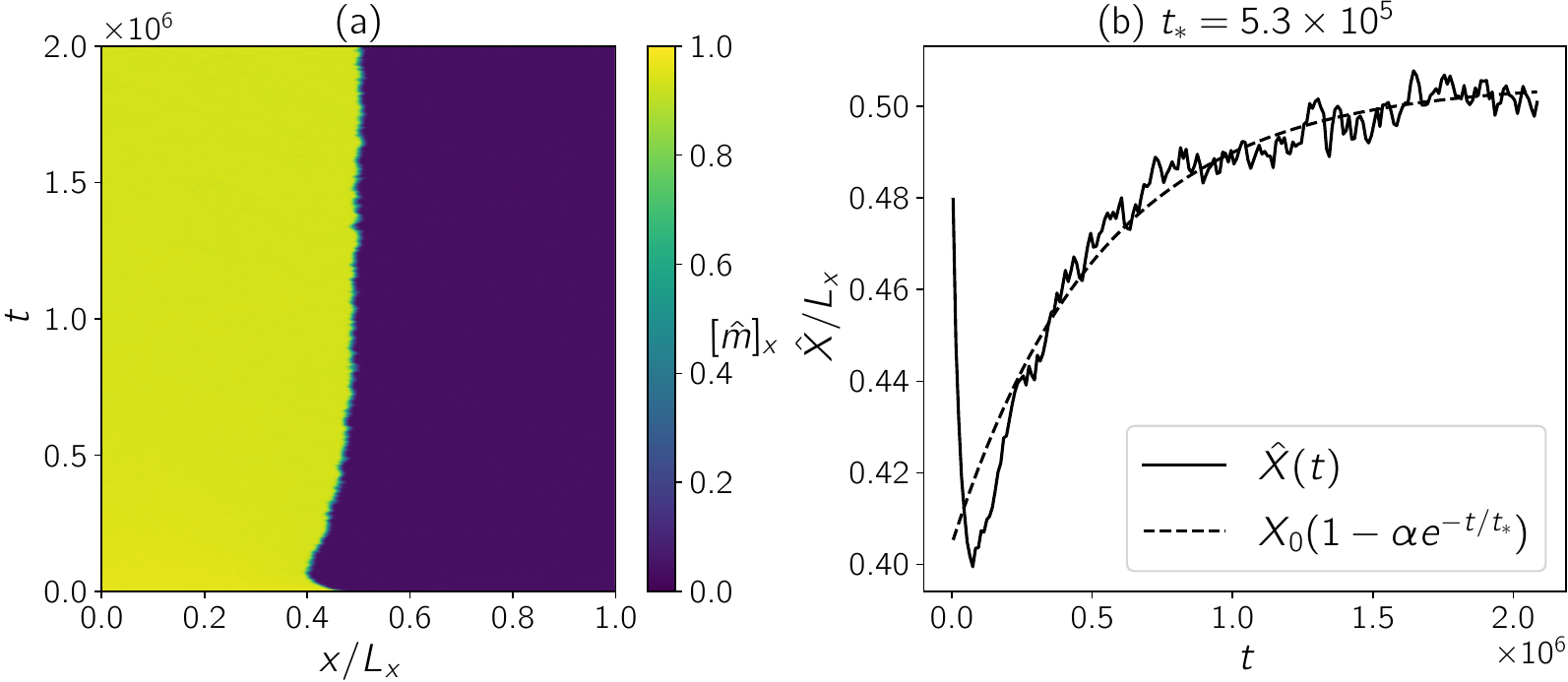}
\caption{
\label{fig:current-dynamics_mc}
(a) Time evolution of the averaged order parameter density
$[\hat m]_x$ according to the Hamiltonian equation \eqref{Hamil} with the boundary condition \eqref{b-con}.
The initial condition is given by a combination of two
equilibrium snapshots with different temperatures $T_1=0.63\Tc$
(left region) and $T_2=1.1\Tc$ (right region). 
(b) Dependence of the interface position $\hat X$ on time $t$.
The dashed lines show the fitting with $X_0 (1 - \alpha e^{-t/t_\ast})$
for $\hat X(t)$. The energy density is set to $E/(L_xL_y)=0.88$.
}
\end{figure}

\begin{figure}[bt]
\centering
\includegraphics[width=0.45\linewidth]{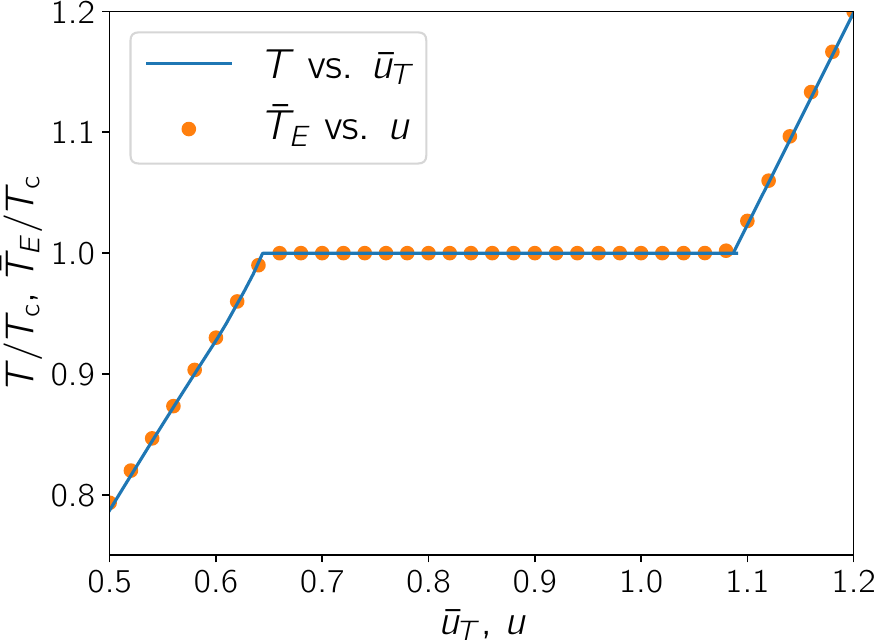}
\caption{
\label{fig:microcanonical_temperature}
(Orange dots) microcanonical temperature $\bar{T}_E$ as a function of energy density $u=E/(L_xL_y)$.
(Blue line) canonical energy density $\bar{u}_T$ as a function of temperature $T$. $T_{\rm c}$ is the transition temperature determined from the canonical simulations.
}
\end{figure}

Figure \ref{fig:current-dynamics_mc} (a) shows an example of
the time evolution  of the one-dimensional order-parameter profile
$[\hat{m}]_x $ as a function of $t$. 
We see that the interface shifts from the initial position and relax to an equilibrium position.
  We expect that the interface dynamics is the slowest in the equilibration
  process.
Then, we estimate the relaxation time $t_\ast$ of
the interface as follows. Let the interface position $\hat X$
be the value of $x$ satisfying $[\hat{m}]_x=0.5$. 
When there are several values of $x$ satisfying $[\hat{m}]_x=0.5$,
  we choose $\hat X$ as their centroid.
In Fig. \ref{fig:current-dynamics_mc} (b), we show the $t$-dependence
of $\hat X$. We define the relaxation time $t_\ast$
by using a functional form $\hat X \simeq X_0(1-\alpha e^{-t/t_\ast})$
in the range 
$0.5\times 10^6 \leq t \leq 2.0 \times 10^6$. We obtain
$t_\ast \simeq 5.3\times 10^5$. 
Based on the result, we define the long-term average of $\hat A(\bv{r},t)$  as 
\begin{align}
\bra {\hat A(\bv{r})}\ket_E
\equiv \frac{1}{t_{\rm f}-t_0}\int_{t_0}^{t_{\rm f}}dt\: \hat A(\bv{r},t),
\label{eq:mcan-expectation}
\end{align}
where we set $t_0=2 \times 10^6$ and $t_{\rm f}=t_0+3.2\times 10^4$. Note that
$t_0\simeq 4t_\ast$, which would be sufficient for equilibration as shown in Fig.  \ref{fig:current-dynamics_mc} (b). 
We thus conjecture that $\sbkt{\hat A(\bv{r})}_E$ is approximately
equal to the expected value  of ${\hat A(\bv{r})}$ with respect to
the microcanonical ensemble
\begin{align}
  \rho_E^\mc(\phi,\pi)=\frac{1}{\Sigma} \delta(E-\mathcal{H}(\phi,\pi)),
  \label{eq:rho-mc}
\end{align}
where  $\Sigma$ is the normalization constant. 

In Fig. \ref{fig:microcanonical_temperature}, we show the numerical results
of 
the microcanonical temperature $\bar{T}_E$ given by
\begin{align}
  \bar T_E = \frac{1}{L_x L_y}\int_D d^2\bv{r}\:
  \bra \hat T(\bv{r}) \ket_E, \quad
  \hat T(\bv{r}) =  \frac{\sum_{a=1}^{q-1}(\pi^a(\bv{r}))^2}{2(q-1)}
\end{align}
as a function of the energy density $u=E/(L_x L_y)$. 
We observe a plateau in a range $0.66 \lesssim u\lesssim 1.09$, 
  which indicates the realization of the phase coexistence.
  We also find good agreement of the data $(u, \bar{T}_E)$
  with another data $(\bar{u}_T, T)$, where $\bar{u}_T$ is the 
  energy density of  an isothermal system with the temperature $T$.
  See Sec. \ref{sec:isothermal} for numerical simulations of the
  isothermal system.

\subsection{Numerical simulations of the Hamiltonian equation
    \eqref{Hamil} with the boundary condition \eqref{b-con-noneq}}

\begin{figure}[bt]
\centering
\includegraphics[width=0.9\linewidth]{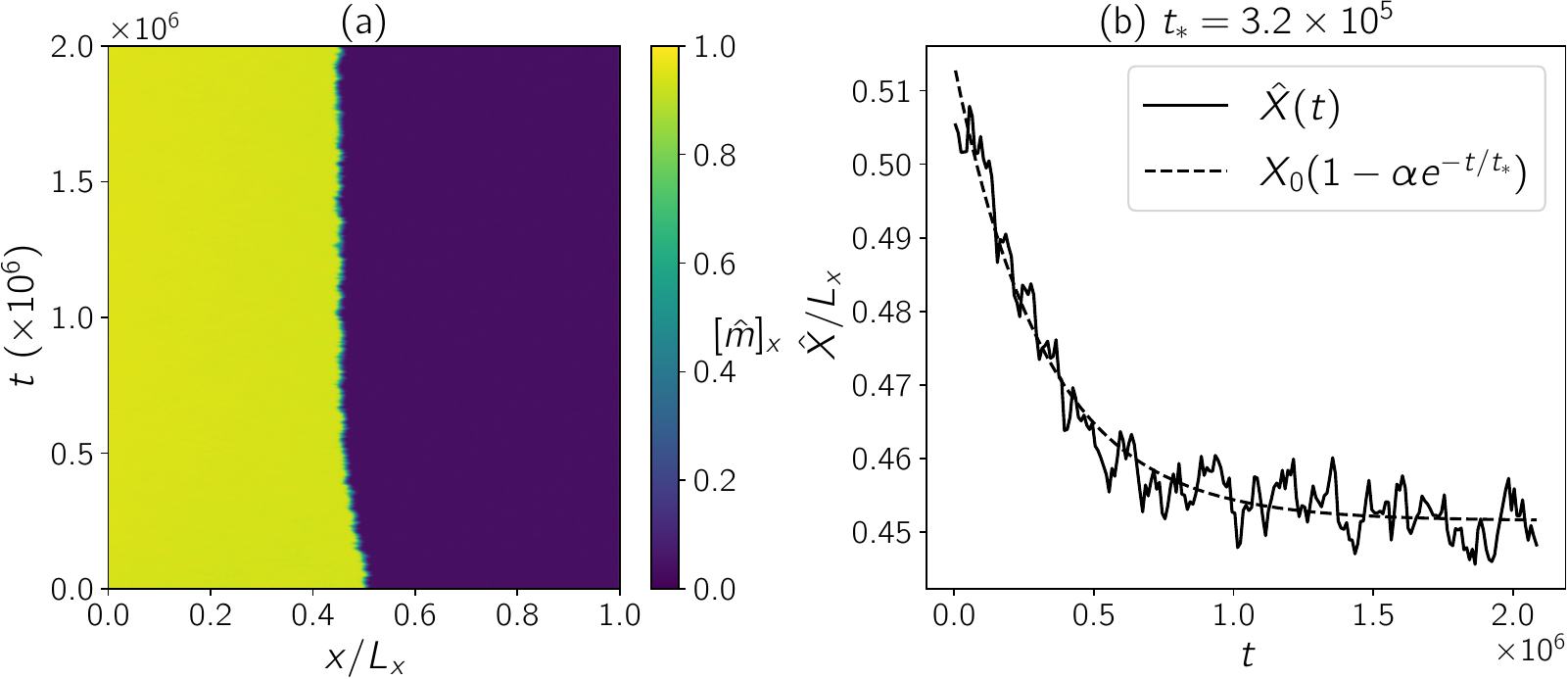} 
\caption{
\label{fig:current-dynamics_ss}
(a) Time evolution  of the averaged order parameter density
$[\hat m]_x$ in the nonequilibrium dynamics under the heat conduction.
(b) Dependence of the interface position $\hat X$ on the time $t$.
The dashed lines show the fitting with $X_0 (1 - \alpha e^{-t/t_\ast})$
for $\hat X(t)$.
The energy density is set to $E/(L_xL_y)=0.88$.
}
\end{figure}

To study the system under the constant heat flux $(JL_y,0)$, 
we solve the Hamiltonian equation \eqref{Hamil} with 
the nonequilibrium boundary condition \eqref{b-con-noneq}.
Since the evolution equation \eqref{Hamil} is unchanged from the previous section, 
we apply the same numerical methods, i.e., spectral collocation and
$4$-th order Strang splitting methods.
The only difference from Sec. \ref{sec:mcan} is the boundary condition,
which is changed from \eqref{b-con} to \eqref{b-con-noneq}.
To apply \eqref{b-con-noneq}, letting $B_{x=0,L_x}$ be the right-hand side in \eqref{b-con-noneq},
\eqref{eq:mcan-Laplacian} should be modified as
\begin{align}
  \nabla^2 \Phi^a(x,y,t) =
  \frac{B_L^a-B_0^a}{L_x}
  -\frac{1}{(N_x-1)N_y}
  \sum_{n_x=0}^{N_x-1}{}^\prime
  \sum_{n_y=-N_y/2+1}^{N_y/2} (k_x^2 + k_y^2)
  \tilde{\Phi}^a(n_x,n_y,t) \cos(k_xx)e^{ik_yy},
\label{eq:Laplacian-mc-neq}
\end{align}
where $B^a_{x=0,L_x}$ is explicitly expressed as 
\begin{align}
  B^a_{x=0,L_x}\equiv
  -\frac{\left. J L_y \Pi^a \right|_{x=0,L_x}}{\int_0^{L_y} dy
    \left. \sum_{b=1}^{q-1} \Pi^{b \ast} \Pi^b \right|_{x=0,L_x}}, \quad
    \Pi^a\equiv\pi^{2a-1}+i\pi^{2a}.
\end{align}
Accordingly, the Fourier
component $\tilde{\Phi}^a(n_x,n_y,t)$
in \eqref{eq:mcan-Laplacian} is replaced by 
\begin{align}
\begin{aligned}
  \tilde{\Phi}^a(n_x,n_y,t) =
  2 \sum_{I=0}^{N_x-1}{}^\prime
  \sum_{J=0}^{N_y-1}\left\{\Phi^a(x,y,t)-
  B_0^a-\frac{(B_L^a-B_0^a)(I\Delta x)^2}{2L}\right\}
  \cos(k_xI\Delta x)e^{-ik_yJ\Delta x}.
\end{aligned}
\end{align}
We denote the solution of the Hamiltonian equation \eqref{Hamil}
with the boundary condition \eqref{b-con-noneq}
as $(\phi,\pi)_{E,J}(\Vec{r},t)$.
We then confirmed that the energy is conserved as $|\bra\mathcal{H}\ket_t-\bra\mathcal{H}\ket_{t=0}|/\bra\mathcal{H}\ket_{t=0}<10^{-8}$.

To observe the effect of the heat flux clearly, we set
the initial condition 
as the final state of the equilibrium simulations performed in the previous section:
$(\phi,\pi)_{E,J}(\Vec{r},0)=(\phi,\pi)_E(\Vec{r},t_{\rm f})$.
Figure \ref{fig:current-dynamics_ss} (a) shows an example of
the  dynamics for the one-dimensional order-parameter profile $[\hat{m}]_x $.
We see that the interface slowly shifts from the initial equilibrium position corresponding to the final position in Fig. \ref{fig:current-dynamics_mc} (a). This clearly shows that the heat flux changes the steady state. 
The relaxation dynamics seems similar to the equilibrium one. 
Then,
we estimate the equilibrating time $t_\ast$ of the interface
in Fig. \ref{fig:current-dynamics_ss} (b) with using  the interface position $\hat X$. 
By fitting $\hat X$ in the range $0 \leq t \leq t_0=2.0 \times 10^6$, we obtain $t_\ast \simeq 3.2\times 10^5$.
Thus, we expect that the system reaches the nonequilibrium steady state before $t_0$ and regard the expected value $\sbkt{\hat  A(\bv{r})}_{E,J}$ in the steady state as
\begin{align}
 \bra {\hat A(\bv{r})}\ket_{E,J}
\equiv \frac{1}{t_{\rm f}-t_0}\int_{t_0}^{t_{\rm f}}dt\: \hat A(\bv{r},t)
\label{eq:mcn-expectation}
\end{align}
with setting $t_{\rm f}=t_0+1.6\times 10^4$.

\subsection{Measurement of heat conductivity}

\begin{figure}[bt]
    \centering
    \includegraphics[width=0.9\linewidth]{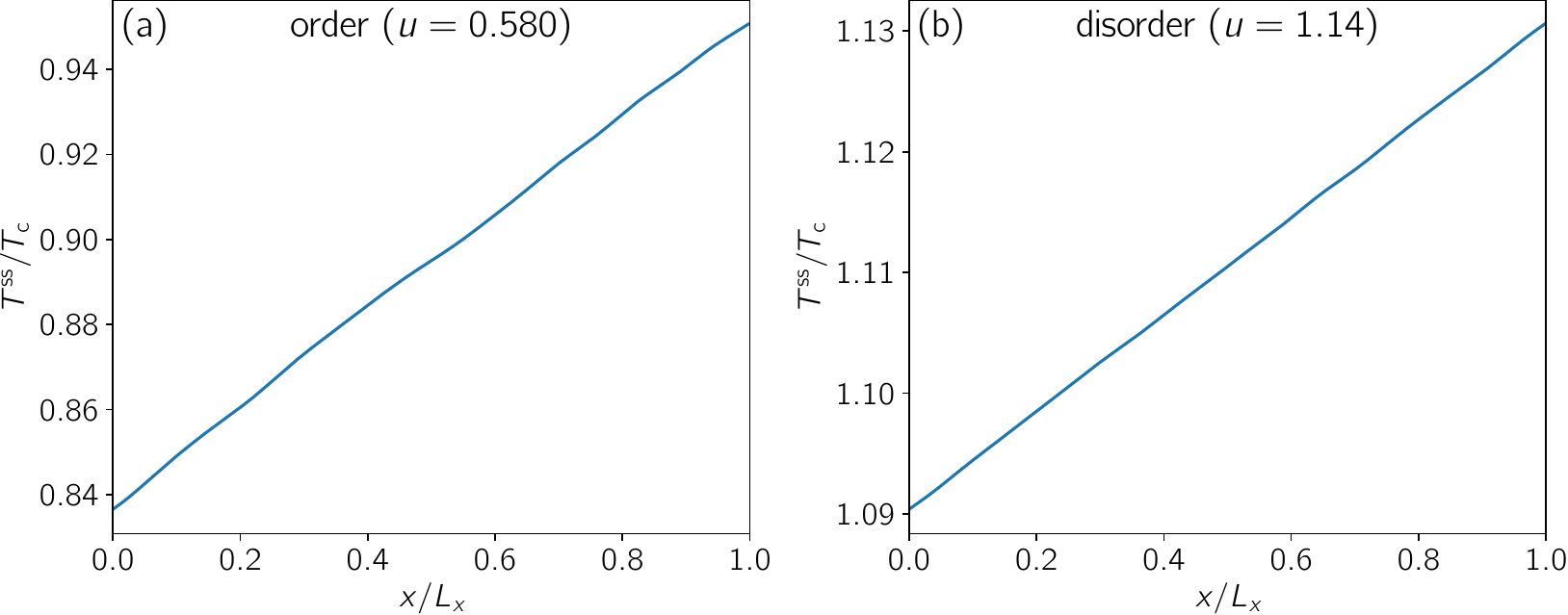}
    \caption{\label{fig:heat_conductivity}
    One dimensional temperature profile $T^{\rm ss}(x)$ with $J=-0.00002$ starting from the initial conditions \eqref{eq:initial_state_for_kappa} with (a) $T=0.9 T_{\rm c}$ (ordered phase with $u=0.580$) and (b) $T=1.1 T_{\rm c}$ (disordered phase with $u=1.14$).
    }
\end{figure}

To numerically evaluate $\theta^{\rm Th}$ in \eqref{e:theta-J},
we have to know the values of heat conductivities $\kappa^{\rm o}$
and $\kappa^{\rm d}$ in ordered and disordered states.
We calculate them  by solving the Hamiltonian equation \eqref{Hamil} with
the nonequilibrium boundary condition \eqref{b-con-noneq}. Here, the initial
condition is given by the final state of \eqref{eq:Ohzeki-Ichiki-under},
that is,  
\begin{align}
    (\phi,\pi_0)_{E,J}(x,y,0)=\begin{cases}
    (\phi_1,\pi_1)_{T}(x,y,t=t_{\rm f}) & 0\leq x<L_x/2, \\
    (\phi_2,\pi_2)_{T}(x-L_x/2,y,t=t_{\rm f}) & L_x/2\leq x<L_x, \\
    (\phi_2,\pi_2)_{T}(0,y,t=t_{\rm f}) & x=L_x.
    \end{cases}
    \label{eq:initial_state_for_kappa} 
\end{align}
Note that the initial condition \eqref{eq:initial_state_for_kappa}
  is an equilibrium configuration at temperature $T$.
  That is, it is an ordered state when $T<\Tc$ and a
disordered state when $T>\Tc$.
We determine the local temperature profile $T^{\rm ss}(x)$ by $\sbkt{\hat [\hat T]_x}_{E,J}$ with  \eqref{eq:mcn-expectation}. 

For the calculation of $\kappa^{\rm o}$, we choose $T=0.9\Tc$ for the initial configuration \eqref{eq:initial_state_for_kappa}.
The corresponding energy density is  $u=0.580$, where the equilibrium system with energy fixed does not show phase coexistence. See Fig. \ref{fig:microcanonical_temperature}. 
As the value of $u$ is sufficiently smaller than the saturation value $E_1/(L_xL_y)$, the local temperature $T^{\rm ss}(x)$ is lower than $\Tc$ everywhere when applying the small constant heat flux $J=-0.00002$. See Fig. \ref{fig:heat_conductivity} (a). 
The local temperature $T^{\rm ss}(x)$ linearly increases from $x=0$ to $x=L_x$, which supports that the value $J=-0.00002$ used in this work is so small for the system to be in the linear-response regime.
We finally determine the heat conductivity by  $\kappa^{\rm o}=-JL_x/(T^\mcn(L_x)-T^\mcn(0))$.

For the calculation of $\kappa^{\rm d}$, we choose $T=1.1\Tc$ for the initial configuration  \eqref{eq:initial_state_for_kappa}, where $u=1.14$.
  As shown in Fig. \ref{fig:heat_conductivity} (b), $T^{\rm ss}(x)$ is higher than $\Tc$ everywhere in $x$ with $J=-0.00002$, and therefore the nonequilibrium system is in the disordered state. The linear temperature profile confirms
that the system is in the linear-response regime. We determine the heat conductivity as $\kappa^{\rm d}=-JL_x/(T^\mcn_{E,J}(L_x)-T^\mcn_{E,J}(0))$.

Finally, we obtain $\kappa^{\rm o}=8.76JL_x/T_{\rm c}$
  and $\kappa^{\rm d}=24.8JL_x/T_{\rm c}$. Using these values,
we plot $\theta^{\rm Th}$ in Fig. \ref{fig:interface} (a).

\subsection{Finite-size effect and roughness effect
  on the interface temperature}

\begin{figure}[htb]
\centering
\begin{minipage}{0.45\linewidth}
\centering
\includegraphics[width=\linewidth]{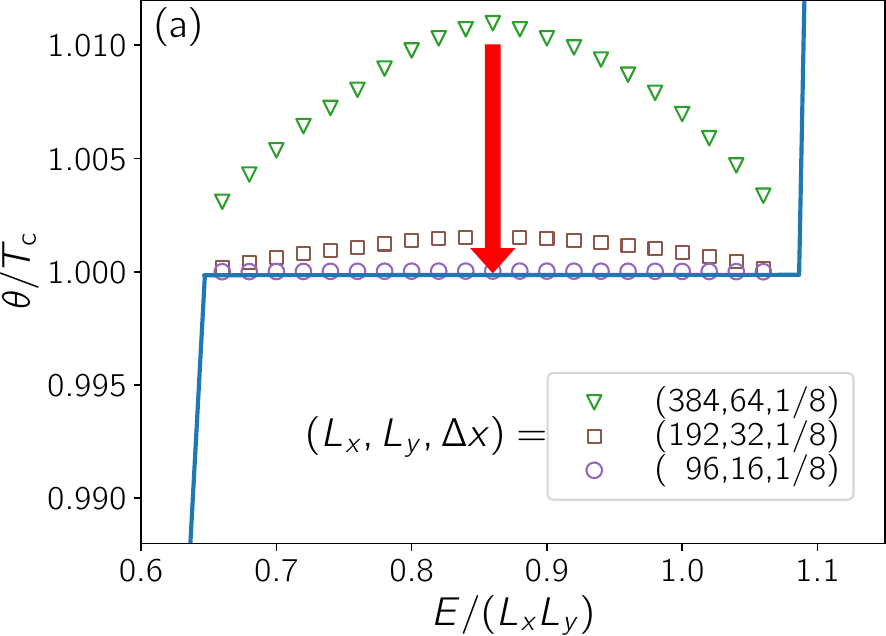}
\end{minipage}
\begin{minipage}{0.45\linewidth}
\centering
\includegraphics[width=\linewidth]{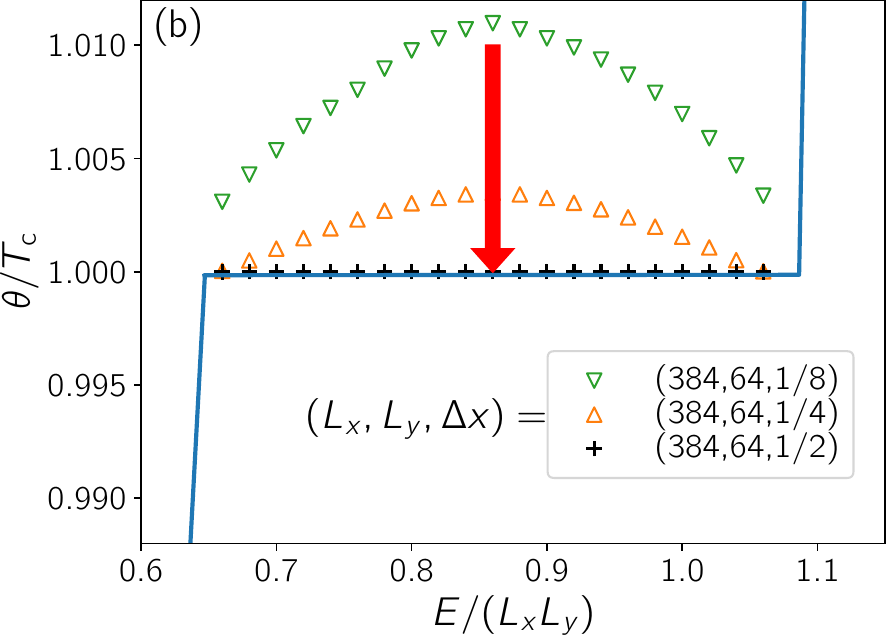}
\end{minipage}
\caption{\label{fig:finite-size-split}
(a) Finite-size dependence of $\theta$ on $(L_x,L_y)=(384,64)$, $(192,32)$ and $(96,16)$ with green inverted triangles, brown squares, and purple circles, respectively.
The grid spacing $\Delta x=1/8$ is fixed.
(b) Roughness dependence of the $\theta$ on $\Delta x=1/8$, $1/4$ and $1/2$ with green inverted triangles, orange triangles, and black crosses, respectively.
The system size $(L_x,L_y)=(384,64)$ is fixed.
In both figures, the blue line indicates the equilibrium functional dependence between $E/(L_xL_y)$ and $T$ obtained by the isothermal simulations explained in Sec. S III.}
\end{figure}

  We separate  the numerical data demonstrated in Fig. \ref{fig:interface}(b) of the main text
  to discuss effects of numerical parameters on the interface
    temperature in the heat conducting systems.
Below, heat flux $J$ is fixed to $J=-0.00002$.

Figure \ref{fig:finite-size-split} (a) shows the dependence of the interface temperature $\theta$ on the system size  with the grid size $\Delta x=1/8$ fixed.
A comparison of the numerical results for $(L_x,L_y)=(384,64)$, $(192,32)$, and $(96,16)$ clarifies the growth of the deviation of  $\theta$ from $\Tc$
for larger system sizes.
When performing numerical simulations in the smallest size $(96,16)$, one might conclude that the local equilibrium explains the heat conducting state well. Indeed, even the smallest system of $(96,16)$ possesses $98,304$ grid points, which may be thought as a large size in numerical simulations. However, our numerical results show that a much larger system size brings a qualitatively different behavior in heat conduction from smaller system sizes.

Next, we show in Fig. \ref{fig:finite-size-split} (b) the dependence on the grid size $\Delta x$ with 
$(L_x, L_y)=(384, 64)$ fixed.
The figure clearly shows that the interface temperature converges to $\Tc$ increasing the roughness of the system as $\Delta x=1/8$, $1/4$ and $1/2$.
The violation of the local equilibrium is hardly visible in
$\Delta x=1/2$.
This fact implies that the violation of the local equilibrium requires not only the large system size but also regularity of fluctuations in a certain short wavelength. 

\section{Determination of the phase diagram in Fig. \ref{fig:simplex} (b)}
\label{sec:isothermal}

In this section, we figure out the calculation for determining the equilibrium phase diagram in Fig. \ref{fig:simplex} (b).
We concentrate on the system at constant temperature $T$.
The equilibrium state is described by the canonical ensemble
\begin{align}
\rho_T^{\can}(\phi,\pi) = \frac{1}{\mathcal{Z}}\exp(-{\cal H}(\phi,\pi)/T),
\label{can}
\end{align}
where $\mathcal{Z}$ is the normalization constant satisfying the condition
\begin{align}
\int {\cal D}\phi {\cal D}\pi \: \rho_T^{\can}(\phi,\pi)=1 .
\end{align}
Since it is not easy to obtain the canonical ensemble
in numerical simulations of the model that exhibits a first-order
transition,  we explain the numerical scheme we adopted.

\subsection{Langevin equation and numerical methods}
\label{sec:langevin}

As an evolution equation consistent with the Hamiltonian equation \eqref{Hamil},
we adopt the Langevin equation
\begin{align}
\begin{aligned}
 \partial_t \phi^a =\var{{\cal H}}{\pi^a}, \quad
 \partial_t \pi^a = -\var{{\cal H}}{\phi^a}-\gamma \pi^a+\xi^a,
\end{aligned}
\label{lan}
\end{align}
where $\xi^a(\bv{r},t)$ is the white Gaussian noise satisfying 
\begin{align}
  \bra \xi^a(\bv{r},t)\xi^b(\bv{r}',t')\ket
  = 2 \gamma T \delta^{ab} \delta(\bv{r}-\bv{r}')  \delta(t-t').
\end{align}
The canonical ensemble \eqref{can}
is given as the stationary distribution of \eqref{lan}.
We set $\gamma = 1$ without loss of generality.

We solve the Langevin equation \eqref{lan} in a periodic box with $(L_x,L_y,\Delta x)=(192,64,1/8)$ consisting of $N_x\times N_y$ grid points with $(N_x,N_y)=(1536,512)$ separated by the lattice spacing $\Delta x$.
To calculate the spatial derivative $\nabla^2 \phi^a$ included in ${\delta \cal H}/\delta \phi^a$,
 we again use the spectral collocation method.
There is a slight difference from \eqref{eq:mcan-Laplacian} because of the difference of the boundary condition.
Concretely, we use
\begin{align}
\begin{aligned}
&
\nabla^2 \Phi^a(x,y,t) = - \frac{1}{N_xN_y}\sum_{n_x=-N_x/2+1}^{N_x/2} \sum_{n_y=-N_y/2+1}^{N_y/2} (k_x^2 + k_y^2) \tilde{\Phi}^a(n_x,n_y,t) e^{i(k_xx+k_yy)}, \\
&
k_x = \frac{2 \pi n_x}{L_x}, \quad
k_y = \frac{2 \pi n_y}{L_y}.
\end{aligned}
\label{eq:Laplacian-can}
\end{align}
Note that $\sum$ denotes the standard summation which is different from 
$\sum^\prime$ defined in \eqref{e:sum-prime}.
The Fourier component $\tilde{\Phi}^a(n_x,n_y,t)$
in \eqref{eq:Laplacian-can} is obtained by
the inverse-Fourier transformation
\begin{align}
\begin{aligned}
\tilde{\Phi}^a(n_x,n_y,t) = \sum_{I=0}^{N_x-1} \sum_{J=0}^{N_y-1} \Phi^a(x,y,t) e^{-i(k_xI+k_yJ)\Delta x} .
\end{aligned}
\end{align}

The $N_x\times N_y$ grid points evolve according to the Langevin equation \eqref{lan}.
To obtain the time evolution of these points,  we use
the 2nd-order
Strang splitting:
\begin{align}
\begin{aligned}
& \phi^a(\Vec{r},t+\Delta t/2) = \phi^a(\Vec{r},t)+\frac{\Delta t}{2}\pi^a(\Vec{r},t), \\
& \pi^a(\Vec{r},t+\Delta t) = \pi^a(\Vec{r},t)-\Delta t\left(\var{{\cal H}}{\phi^a}(\Vec{r},t+\Delta t/2)+\pi^a(\Vec{r},t)\right) +\sqrt{2T(\Delta x)^2(\Delta t)}\Theta^a(\Vec{r},t),\\
& \phi^a(\Vec{r},t+\Delta t) = \phi^a(\Vec{r},t+\Delta t/2)+\frac{\Delta t}{2}\pi^a(\Vec{r},t+\Delta t),
\end{aligned}
\label{eq:discrete-lan}
\end{align}
where $\Delta t=1/4096$ and $\Theta^a(\Vec{x},t)$ are the random numbers generated by the standard normal distribution.
Below, we denote the solution of the Langevin equation
\eqref{lan} as $(\phi,\pi)_{T}(\Vec{r},t)$.

\subsection{Metastable and equilibrium states}

Compared with the Hamiltonian equation \eqref{Hamil} investigated in Sec. \ref{sec:mcan}, the Langevin equation \eqref{lan} shows a faster relaxation because of no energy conservation. Since a typical relaxation time of the Langevin equation is independent of system sizes, it is  four orders of magnitude shorter
than that for the Hamiltonian equation where the relaxation time is
proportional to $L_x^2$.
However, near the first-order transition, the system may relax to the metastable state rather than the equilibrium state. Once the system is trapped in the metastable state, it takes significant time to get out of it. Since the observed state depends on the initial condition, the system shows hysteresis.

Throughout the following sections,
  we fix the transient time as $t_0=1.0\times 10^3$, 
  and observe states from $t_0$ to $t_{\rm f}=t_0+1.6\times 10^4$.
The time $t_0$ is much longer than the relaxation time to the equilibrium or metastable states, however, much shorter than the first passage time to get out of the metastable state.
The observation period $t_{\rm f}-t_0$ is so long that
the finite time fluctuations can be ignored.
In such a time scale, we cannot distinguish the metastable state from the equilibrium stats.
Thus, we first examine the hysteresis and then proceed to
  the introduction of
  a numerical scheme for providing the canonical ensemble.
Our aim is to determine the equilibrium energy density of the system
\begin{align}
\bar{u}_T= \frac{1}{L_x L_y} \int {\cal D}\phi {\cal D}\pi \: \rho_T^{\can}(\phi,\pi) {\cal H}(\phi,\pi) ,
\label{eq:expectation-can-u}
\end{align}
and the order parameter of the system
\begin{align}
\bar{m}_T= \frac{1}{L_x L_y} \int_D d^2\bv{r}\:  \int {\cal D}\phi {\cal D}\pi \: \rho_T^{\can}(\phi,\pi) {\hat m}(\bv{r}) ,
\label{eq:expectation-can-m}
\end{align}
which provides a phase diagram. 

Below, we examine the equilibrium and metastable states over a wide range of $T$ in $[T_{\rm min}, T_{\rm max}]$, where $T_{\min}= 0.0015$ and $T_{\max}= 0.3$.
Precisely, we examine $201$ temperature points, 
\begin{align}
T_k=T_{\rm min}+k\Delta T, \quad (k=0,1,\cdots, 200),
\end{align}
where $\Delta T\equiv (T_{\rm max}-T_{\rm min})/200$.
Using the numerical data at various temperatures, we specify the equilibrium transition temperature $\Tc$ in Sec. \ref{e:sec:equilibrium}.
The results of the equilibrium simulations are shown in Figs. \ref{fig:canonical} (a) and (b), where $\bar{u}_T$ and $\bar m_T$
exhibits the discontinuous jump at $T =\Tc \simeq 0.150$.
This indicates that the first-order transition occurs at $T =\Tc$.
Figure \ref{fig:canonical} (a) is the same as that
  displayed in Fig. \ref{fig:simplex} (b).

\subsubsection{Heating experiment}
\label{sec:heating}

We start with the initial condition $(\phi,\pi)_{T_{\rm min}}(\Vec{r},0)=(\mu_1,0)$ at $T=T_{\rm min}$.
Note that the direction of the symmetry breaking is specified by this initial condition in the following heating  experiment.
We numerically solve the Langevin equation \eqref{lan}
up to $t=t_{\rm f}$  at $T_{\rm min}=T_{k=0}$, and calculate
  the time average of observables over the time interval $t_{\rm f}-t_0$.
That is, for 
\begin{align}
&  \bar{u}_{T_k}^{\rm heat}\equiv \frac{1}{L_x L_y}\frac{1}{t_{\rm f}-t_0}  \int_{t_0}^{t_{\rm f}}  dt\: {\cal H}(\phi(t),\pi(t)),\label{eq:can-expectation-u-heat}
\\
&  \bar{m}_{T_k}^{\rm heat}\equiv \frac{1}{L_x L_y}\int_D d^2\bv{r}\:  \frac{1}{t_{\rm f}-t_0}  \int_{t_0}^{t_{\rm f}}  dt\: {\hat m}(\bv{r},t),
\label{eq:can-expectation-m-heat}
\end{align}
we obtain $(\bar{u}_{T_0}^{\rm heat}, \bar{m}_{T_0}^{\rm heat})$.
We then increment the temperature from $T_0$ to $T_1$
and reset the time to be $t=0$. The initial condition
  at $T=T_1$ is equal to the final state at $T_0$, i.e.,
  $  (\phi,\pi)_{T_1}(\Vec{r},0)=(\phi,\pi)_{T_{0}}(\Vec{r},t_{\rm f})$. 
  We then numerically solve the Langevin equation \eqref{lan}
  up to $t=t_{\rm f}$. We repeat the same procedure changing the
  temperature from $T_{k-1}$ to
$T_k$ and setting   $  (\phi,\pi)_{T_k}(\Vec{r},0)
=(\phi,\pi)_{T_{k-1}}(\Vec{r},t_{\rm f})$, where $1 \le k \le 200$. 
  We plot the temperature dependence of $\bar{u}_{T_k}^{\rm heat}$ and
  $ \bar{m}_{T_k}^{\rm heat}$ 
as displayed in Figs.  \ref{fig:canonical} (c) and (d).
At $T\simeq0.166$, the system suddenly enters the disordered
state with $\bar{m}_T^{\rm heat}\simeq 0$.
 At the same temperature, $\bar{u}_T^{\rm heat}$ also
shows a discontinuous jump.

\subsubsection{Cooling experiment}
\label{sec:cooling}

\begin{figure}[bt]
\centering
\includegraphics[width=0.8\linewidth]{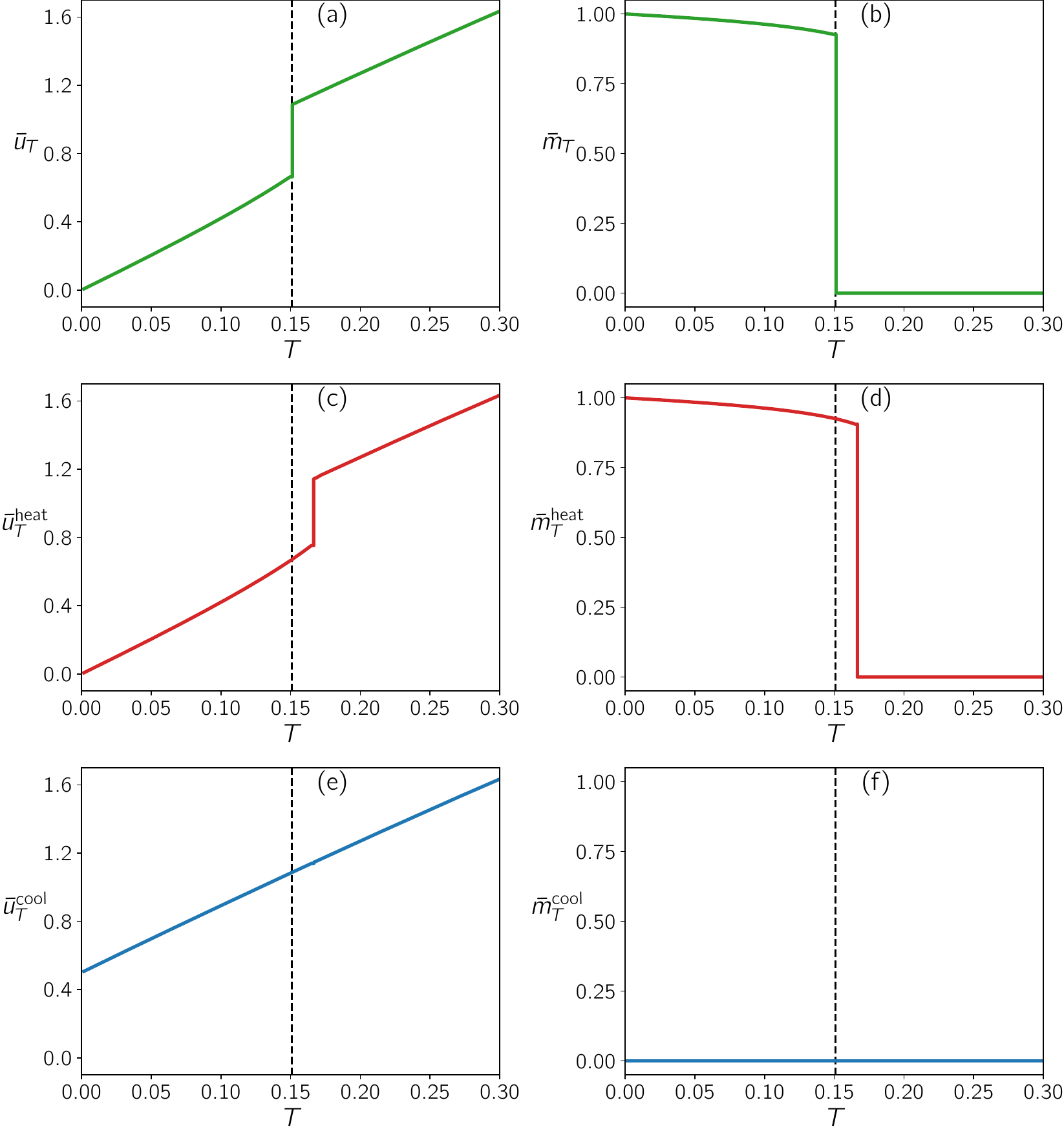}
\caption{
\label{fig:canonical}
Temperature dependence of the energy density $\bar{u}_T$ (panels (a), (c), and (e)) and the order parameter $\bar{m}_T$ (panels (b), (d), and (f)) as the function of $T$ for the equilibrium with the Langevin equation \eqref{eq:Ohzeki-Ichiki-under} (panels (a) and (b)), the heating experiment with the Langevin equation \eqref{lan} (panels (c) and (d)), and the cooling experiment with the Langevin equation \eqref{lan} (panels (e) and (f)).
}
\end{figure}

  We start with an initial state which is obtained as
the final state of the heating experiment.
We then decrease the temperature from $T_k$ to $T_{k-1}$,
where $k$ changes from $k=200$ to $k=1$. From each $T_k$, the time
is reset as $t=0$, and the initial condition at $T=T_{k-1}$ is
$(\phi,\pi)_{T_{k-1}}(\Vec{r},0)=(\phi,\pi)_{T_{k}}(\Vec{r},t_{\rm f})$
in the cooling experiment.
Similarly to the heating experiment, we define
\begin{align}
&  \bar{u}_{T_k}^{\rm cool}\equiv \frac{1}{L_x L_y}\frac{1}{t_{\rm f}-t_0}  \int_{t_0}^{t_{\rm f}}  dt\: {\cal H}(\phi(t),\pi(t)),\label{eq:can-expectation-u-cool}\\
&  \bar{m}_{T_k}^{\rm cool}\equiv \frac{1}{L_x L_y}\int_D d^2\bv{r}\:  \frac{1}{t_{\rm f}-t_0}  \int_{t_0}^{t_{\rm f}}  dt\: {\hat m}(\bv{r},t).
\label{eq:can-expectation-m-cool}
\end{align}
At a glance, \eqref{eq:can-expectation-u-cool} and \eqref{eq:can-expectation-m-cool} are the same as  \eqref{eq:can-expectation-u-heat}
and \eqref{eq:can-expectation-m-heat}.
The only but significant difference is the initial condition for each $T_k$.
In the cooling experiment,
we obtain Figs.  \ref{fig:canonical} (e) and (f).
They are obviously different from Figs.  \ref{fig:canonical} (c) and (d).
Through the cooling experiment, the system always shows the
disordered state with $\bar{m}_T\simeq 0$ and the symmetry
breaking never occurs, which suggests that the supercooled
metastable disordered state can exist even close to the zero temperature.
These results indicate the existence of a strong hysteresis in this system.


\subsubsection{Equilibrium values}
\label{e:sec:equilibrium}

Now, we determine which quantity 
$(\bar{u}_T^{\rm heat}, \bar{m}_T^{\rm heat})$ or
$(\bar{u}_T^{\rm cool}, \bar{m}_T^{\rm cool})$
corresponds to the equilibrium value.
Here,  we use the equilibration method developed by Ohzeki and Ichiki
\cite{Ohzeki-Ichiki2015-sm}.  We consider a duplicated system 
comprising $(\phi_{\rm A},\pi_{\rm A})$ and $(\phi_{\rm B},\pi_{\rm B})$, each of which is defined
in the periodic box with $(L_x, L_y) = (192,64)$.
The duplicated system evolves by
\begin{align}
\begin{aligned}
    &
    \partial_t\phi^a_{\rm A}=\var{{\cal H}(\phi_{\rm A},\pi_{\rm A})}{\pi^a_{\rm A}}-\var{{\cal H}(\phi_{\rm B},\pi_{\rm B})}{\pi^a_{\rm B}}, \quad
    \partial_t\pi^a_{\rm A}=-\var{{\cal H}(\phi_{\rm A},\pi_{\rm A})}{\phi^a_{\rm A}}-\var{{\cal H}(\phi_{\rm B},\pi_{\rm B})}{\phi^a_{\rm B}}-\pi^a_{\rm A}+\xi^a_{\rm A}, \\
    &
    \partial_t\phi^a_{\rm B}=\var{{\cal H}(\phi_{\rm B},\pi_{\rm B})}{\pi^a_{\rm B}}+\var{{\cal H}(\phi_{\rm A},\pi_{\rm A})}{\pi^a_{\rm A}}, \quad
    \partial_t\pi^a_{\rm B}=-\var{{\cal H}(\phi_{\rm B},\pi_{\rm B})}{\phi^a_{\rm B}}+\var{{\cal H}(\phi_{\rm A},\pi_{\rm A})}{\phi^a_{\rm A}}-\pi^a_{\rm B}+\xi^a_{\rm B},
\end{aligned}
\label{eq:Ohzeki-Ichiki-under}
\end{align}
where $\xi_i^a(\Vec{r},t)$ satisfies
\begin{align}
  \bra \xi_i^a(\bv{r},t)\xi_j^b(\bv{r}',t')\ket
  = 2 \gamma T \delta^{ab} \delta_{ij} \delta(\bv{r}-\bv{r}')  \delta(t-t').
\end{align}
Writing down the Fokker-Planck equation, we can confirm that the stationary distribution of 
\eqref{eq:Ohzeki-Ichiki-under}  coincides with the canonical distribution \eqref{can}.
It is also proved that the relaxation time of \eqref{eq:Ohzeki-Ichiki-under} to the true equilibrium is shorter than that of the original system \eqref{lan} \cite{Ohzeki-Ichiki2015-sm}.

We solve \eqref{eq:Ohzeki-Ichiki-under}
by discretizing the system with the lattice spacing $\Delta x=1/8$ and using the same methods as those for
the original Langevin equation \eqref{lan}.
The initial condition at the temperature $T$ is given as  a couple of the final states obtained in the heating and cooling experiments,
\begin{align}
(\phi_{\rm A},\pi_{\rm A})_T(\Vec{r},0)&=(\phi,\pi)_T(\Vec{r},t_{\rm f})^{\rm heat}, \\ 
(\phi_{\rm B},\pi_{\rm B})_T(\Vec{r},0)&=(\phi,\pi)_T(\Vec{r},t_{\rm f})^{\rm cool},
\end{align}
where $(\phi,\pi)_T(\Vec{r},t_{\rm f})^{\rm heat}$ and $(\phi,\pi)_T(\Vec{r},t_{\rm f})^{\rm cool}$ are the final states at $T$ calculated in Sec. \ref{sec:heating} and Sec. \ref{sec:cooling}, respectively.
Let $t_0$ and $t_{\rm f}$ be the same value used in numerical simulations
of the original Langevin equation \eqref{lan}. 
We observe that the subsystems A and B
become to exhibit the same statistical state
until $t=t_0$.
We thus expect that the obtained states
$(\phi_{\rm A},\pi_{\rm A})_T(\Vec{r},t)$ and
$(\phi_{\rm B},\pi_{\rm B})_T(\Vec{r},t)$ for $t \ge t_0$ obey 
the respective canonical distribution 
$\rho_T^\can(\phi_{\rm A},\pi_{\rm A})$ and
$\rho_T^\can(\phi_{\rm B},\pi_{\rm B})$ defined in \eqref{can}.
The equilibrium values $\bar{u}_T$ and $\bar{m}_T$ defined
in \eqref{eq:expectation-can-u} and \eqref{eq:expectation-can-m}
are then given by time averages in the interval $t_{\rm f}-t_0$.
Explicitly, we calculate them as
\begin{align}
& \bar{u}_{T} =  \frac{1}{L_x L_y}\frac{1}{t_{\rm f}-t_0}  \int_{t_0}^{t_{\rm f}}  dt\: \frac{ {\cal H}(\phi_{\rm A}(t),\pi_{\rm A}(t)) + {\cal H}(\phi_{\rm B}(t),\pi_{\rm B}(t))}{2},   \label{eq:ensemble-Ichiki-Ohzeki-u}\\
&  \bar{m}_{T} =
\frac{1}{L_x L_y}\int_D d^2\bv{r}\: \frac{1}{t_{\rm f}-t_0}  \int_{t_0}^{t_{\rm f}} dt\:
\sum_{a=1}^{q-1}\frac{\phi_{\rm A}^a(\bv{r})+\phi_{\rm B}^a(\bv{r})}{2}\mu_1^a.
  \label{eq:ensemble-Ichiki-Ohzeki-m}
\end{align}
Since the duplicated system has $(\phi_{\rm A},\pi_{\rm A})$ and $(\phi_{\rm B},\pi_{\rm B})$, we take the arithmetic means of the two in \eqref{eq:ensemble-Ichiki-Ohzeki-u} and \eqref{eq:ensemble-Ichiki-Ohzeki-m}.
Figures \ref{fig:canonical} (a) and (b) show $\bar{u}_T$
and $\bar{m}_T$ obtained by \eqref{eq:ensemble-Ichiki-Ohzeki-u} and \eqref{eq:ensemble-Ichiki-Ohzeki-m}, respectively.

In Fig. \ref{fig:microcanonical_temperature}, we also show the
data of Fig. \ref{fig:canonical} (a) in addition to
the microcanonical temperature $\bar{T}_E$ as a function of $u=E/(L_xL_y)$.
The both data show quantitatively the same behavior including the states
with the interface at $T=\bar{T}_E=T_{\rm c}$, which shows the
equivalence of thermodynamic ensembles \eqref{eq:rho-mc} and \eqref{can}.
This result also demonstrates the validity of the numerical method
generating the canonical ensemble \eqref{can}.


\begin{thebibliography}{99}
\bibitem{fn:LR}
{This value was chosen so that the system is in the linear response
regime in $J$. For example, linear temperature profiles are 
observed in the ordered and disordered phases \cite{SM}. }
\end{thebibliography}
\end{document}